\documentclass[aps,pra,twocolumn,floatfix]{revtex4-2}
\usepackage{graphicx}
\usepackage{amssymb}
\usepackage{amsmath}
\usepackage{bm}
\usepackage{color}
\usepackage{epstopdf}

\usepackage[normalem]{ulem} 

\begin{document}

\title{Efficient First-Principles Approach with a Pseudohybrid Density Functional for Extended Hubbard Interactions}

\author{Sang-Hoon Lee}
\email{E-mail: lshgoal@kias.re.kr}
\author{Young-Woo Son}
\email{E-mail: hand@kias.re.kr}
\affiliation{Korea Institute for Advanced Study, Seoul 02455, Korea}

\begin{abstract}
For massive database-driven materials research, 
there are increasing demands for both fast and accurate quantum mechanical computational tools. 
Contemporary density functional theory (DFT) methods can be fast sacrificing their accuracy or 
be precise consuming a significant amount of resources. 
Here, to overcome such a problem,
we present a DFT method that exploits self-consistent determinations of 
the on-site and inter-site Hubbard interactions ($U$ and $V$) simultaneously
and obtain band gaps of diverse materials 
in the accuracy of $GW$ method at a standard DFT computational cost.
To achieve self-consistent evaluation of $U$ and $V$, 
we adapt a recently proposed Agapito-Curtarolo-Buongiorno Nardelli
pseudohybrid functional for $U$ to implement a new density functional of $V$. 
This method is found to be appropriate for considering various interactions  
such as local Coulomb repulsion, covalent hybridization and their coexistence. 
We also obtained good agreements between computed and measured band gaps of low dimensional systems, 
thus meriting the new approach for large-scale as well as high throughput calculations for various bulk and nanoscale materials
with higher accuracy.
\end{abstract}

\maketitle

\section{Introduction\label{intro}}

Theoretical and computational methods based on the density functional theory (DFT)~\cite{HK,KS} 
have been indispensable tools in understanding physical properties of real materials~\cite{Jones2015RMP}. 
Although they fail quantitatively and sometimes qualitatively in calculating band gaps~\cite{Jones2015RMP}
with the local density approximation (LDA)~\cite{KS}
or the generalized-gradient approximation (GGA)~\cite{PBE},
they are only currently available methods without significant computational costs 
to offer fully quantum mechanical computational results
for diverse phenomena involved with many thousands of electrons~\cite{Jones2015RMP,Kummel2008RMP}.
Thus, regardless of such shortcomings, the DFT-based approaches prevail
in data-driven materials researches~\cite{Curtarolo2013NatMat} 
spanning various areas such as energy materials~\cite{Greeley2006NatMat,Wang2011PRX,Yu2012PRL},
electronic applications~\cite{Armiento2011PRB,Hautier2013NatComm,Yim2015NPG},
low dimensional crystals~\cite{Mounet2018NatNano}
and topological materials~\cite{Yang2012NatMat,Zhang2019Nature,Vergniory2019Nature,Tang2019Nature}.
These databases with improved accuracy will be of great benefit 
in advancing future technology.

To build high-quality materials databases, 
it is vital to improve the accuracy of DFT-based methods.
Several methods beyond LDA and GGA have been suggested so far. 
The local Coulomb repulsion $U$ was introduced in DFT+$U$ 
to compensate the overdelocalization of $d$- or $f$-electrons
in LDA or GGA~\cite{Anisimov1991PRB,Anisimov1997JPC}.
Beyond static correlation effectively treated in DFT+$U$, 
DFT with the dynamical mean-field theory~\cite{DMFT1,DMFT2,Kotliar2006RMP} 
has been used for strongly interacting materials.
The quasiparticle energy of semiconductors can be obtained accurately
with the $GW$ approximations~\cite{Hedin1965PR,Hybertsen1986PRB,Shishkin2007PRL}.
Hybrid functionals such as HSE ~\cite{Heyd2003JCP, Janesko2009PCCP} 
and LDA with the modified Becke-Johnson exchange potential (mBJLDA)~\cite{Tran2009PRL} 
are also popular.
However, all the methods above except DFT+$U$ and mBJLDA involves intensive computations discouraging their use in data-driven researches. 
Due to some limitations in the latter~\cite{Koller2011PRB},
we will focus on improving the former for high-throughput calculations. 

Two aspects in the DFT+$U$ formalism are important in
obtaining accurate band gaps {\it ab initio}.
First, the on-site Hubbard $U$ needs to be estimated
self-consistently~\cite{Kulik2006PRL} and 
various methods for this purpose~\cite{Coco2005PRB,Miyake2008PRB,
Aichhorn2009PRB,Miyake2009PRB,Kulik2006PRL,Mosey2007PRB,Mosey2008JCP,Agapito2015PRX}
have been suggested.
Among them, the direct evaluation from the Hartree-Fock (HF) formalism~\cite{Mosey2007PRB,Mosey2008JCP,Agapito2015PRX} 
is relevant here since other methods involve additional expensive calculations.
A recent proposal by Agapito-Curtarolo-Buongiorno Nardelli 
(ACBN0)~\cite{Agapito2015PRX} allows a direct self-consistent evaluation of $U$.
They demonstrated improved agreements with experiments
with a negligible increase in computational cost~\cite{Agapito2015PRX,Rubio2017PRB}.
Second, the inter-site Hubbard $V$ between the localized orbital of interest in DFT+$U$
and its neighboring orbitals also need to be considered properly 
because it could lead to better descriptions of electronic structures
of some solids~\cite{Anisimov1996PRB,Cococcioni2010JPC,Kulik2011JCP}.
Moreover, DFT+$U$ hardly improves LDA and GGA gaps 
of simple semiconductors such as Si 
while DFT+$U$ with $V$ does~\cite{Cococcioni2010JPC}.
Therefore, by combining these progresses, we may obtain an
efficient large-scale and high-throughput computational tool for materials researches.

In this paper, we extend the ACBN0 functional for DFT+$U$ ~\cite{Agapito2015PRX} to implement a 
new density functional for the inter-site Coulomb interaction of $V$.
With this, we achieve excellent agreements between the self-consistent {\it ab initio} band gaps 
of diverse semiconductors and insulators and those from experiments. 
The band gaps comparable to those from 
$GW$ methods~\cite{Hybertsen1986PRB,Shishkin2007PRL}
can be obtained within the standard DFT-GGA computational time.
Moreover, for low dimensional systems in which 
the screening of Coulomb interaction varies significantly,
the new method can also compute the accurate band gaps of few layers black phosphorous
and Si(111)-(2$\times$1) surface, respectively,
demonstrating its flexibility on structural and dimensional variations.
Considering recent explosive expansion of data-driven materials 
researches using the DFT~\cite{Curtarolo2013NatMat, Greeley2006NatMat,Wang2011PRX,Yu2012PRL,Armiento2011PRB,Hautier2013NatComm,Yim2015NPG,Mounet2018NatNano,Yang2012NatMat,Zhang2019Nature,Vergniory2019Nature,Tang2019Nature}, 
the improved accuracy in DFT computations is of great importance 
in constructing useful and reliable databases of materials. 
Thus, we expect that this new approach could accelerate efficient
high-throughput calculations with better accuracy for materials discovery.

This paper is organized as follows. We first introduce our formalism of the ACBN0-like functional for the
intersite Hubbard interactions in Sec.~\ref{formalism}. 
Then, using the new method described in Sec.~\ref{method}, 
we present our computational results of energy band gaps of various 
three-dimensional solids in Sec.~\ref{3D} and those of low dimensional systems in Sec.~\ref{2D}. 
Finally, we discuss several aspects of the new functional and conclude in Sec.~\ref{discussion}.

\section{Formalism\label{formalism}}

Let us first consider mean-field (MF) energy of the Coulomb interaction between electrons in
a pair of atoms $I$ and $J$ with the HF approximation,
\begin{eqnarray}
E_\textrm{MF} &=&\frac{1}{2}\sum_{IJ}
\sum_{ij} \sum_{\sigma\sigma'}
\langle \phi^I_i \phi^J_j | V_{ee} | \phi^I_i \phi^J_j \rangle \nonumber \\
&\times&    
\left( n^{II\sigma}_{ii} n^{JJ\sigma'}_{jj}
       -\delta_{\sigma \sigma'} n^{IJ\sigma}_{ij} n^{JI\sigma'}_{ji}
\right).
\label{Eq:HF}
\end{eqnarray}
In the abbreviated representation of pairwise HF energy in Eq.~(\ref{Eq:HF}), the general occupation matrix is written as
\begin{eqnarray}
n^{IJ\sigma}_{ij}&\equiv& n^{I,n,l,J,n',l',\sigma}_{ij}\nonumber \\
                            &=& \sum_{m{\bf k}}w_{\bf k}f_{m\bf k}
\langle \psi_{m{\bf k}}^\sigma |\phi^{I,n,l}_i \rangle \langle \phi^{J,n',l'}_j |\psi_{m{\bf k}}^\sigma\rangle,
\label{Eq:GO}
\end{eqnarray}
where $f_{m\bf k}$ is the Fermi-Dirac function of 
the Bloch state $|\psi_{m\bf k}^\sigma\rangle$ with a spin $\sigma$
of the $m$-th band at a momentum ${\bf k}$
and
$w_{m\bf k}$ is the ${\bf k}$-grid weight.
The L{\"o}wdin orthonormalized atomic wavefunction ($|\phi^{I,n,l}_i\rangle$) is used
as a projector for the localized atomic orbital having the principal, azimuthal, and angular quantum numbers of $n$, $l$, and $i$, respectivley
at an atom $I$. 
We will use a brief notation for atom $I$ representing a specific principal and azimuthal 
quantum numbers of $n$ and $l$ of the $I$-th atomic element in a solid hereafter.  
We also note that the diagonal terms in Eq.~(\ref{Eq:GO}) are the usual on-site occupations for DFT+$U$.
In Eq.~(\ref{Eq:HF}), we neglect other small pairwise interactions, {\it e.g.,} the cross charge exchanges 
between neighbors~\cite{Cococcioni2010JPC},
and discuss their effects in Appendix A and in the Table~\ref{Table_A1}.

Assuming the effective interactions of $\langle V_{ee}\rangle$ in Eq.~(\ref{Eq:HF})
are all equal to their atomic average~\cite{Cococcioni2010JPC},
the rotationally invariant or angular momentum averaged form of $E_\text{MF}$ can be written as 
$E_\text{MF}=E_\textrm{Hub}=E_U+E_V$ 
where $E_U$ is for the case of $I=J$ and $E_V$ for $I\neq J$.
$E_U$ is the well-known energy functional for $U$ suggested by Dudarev {\it et al.}~\cite{Dudarev1998PRB}.
For $I\neq J$ case where atoms $I$ and $J$ locate at {\it different} positions, respectively,
\begin{equation}
E_V =\frac{1}{2}\sum_{\{I,J\}} 
\sum_{ij} \sum_{\sigma \sigma'} V^{IJ} 
\left(
n^{II\sigma}_{ii} n^{JJ\sigma'}_{jj} 
- \delta_{\sigma \sigma'} n^{IJ\sigma}_{ij} n^{JI\sigma'}_{ji}
\right),
\label{Eq:EVIJ}
\end{equation}
where the $\{I,J\}$ indicates the summation 
for pairs of atoms $I$ and $J$ of which interatomic distance of $d_{IJ}$ 
is less than a given cutoff.
In Eq.~(\ref{Eq:EVIJ}), $V^{IJ}$ is the inter-site Hubbard interaction
for the pair and will be determined based on the method of ACBN0~\cite{Agapito2015PRX}.

To obtain a functional form of $V^{IJ}$, as is discussed for $U$ in Ref.~\onlinecite{Agapito2015PRX},
we also follow a central ansatz by Mosey {\it et al.}~\cite{Mosey2007PRB,Mosey2008JCP}
that leads to a ``renormalized" occupation number for the pair such as
\begin{eqnarray}
  N^{IJ\sigma}_{\psi_{m{\bf k}}} &\equiv& 
  N^{I,n,l,J,n',l'\sigma}_{\psi_{m{\bf k}}} \nonumber \\
  &= & 
   \sum_{\{I\}}\sum_{i}
  \langle \psi_{m{\bf k}}^\sigma |\phi^{I,n,l}_i \rangle \langle \phi^{I,n,l}_i |\psi_{m{\bf k}}^\sigma\rangle \nonumber \\
  &+&  \sum_{\{J\}}\sum_{j}
   \langle \psi_{m{\bf k}}^\sigma |\phi^{J,n',l'}_j \rangle \langle \phi^{J,n',l'}_j |\psi_{m{\bf k}}^\sigma\rangle,
\label{Eq:RON}
\end{eqnarray}
where the summation performs for all the orbitals of 
a type $I$ atom with quantum number $n$ and $l$ (denoted by $\{I\}$ in Eq.~\ref{Eq:RON})
and for a type $J$ atom with $n'$ and $l'$ (denoted by $\{J\}$) in a given unit cell, respectively.
Here we note that the sum is only obtained for all the given pairs within a specific distance.
Therefore, we can obtain the ACBN0-like functional for $V^{IJ}$ that effectively accounts
the screening in the bond region between the pair, e.g., the interaction between
$l$-th orbital of $I$ atom and $l'$-th orbital of $J$ atom.
Corresponding to the ACBN0 functional~\cite{Agapito2015PRX} for $U$, 
we can replace $n^{IJ\sigma}_{ij}$ in Eq.~(\ref{Eq:HF})
by a renormalized density matrix for the pair, 
\begin{equation}
P^{IJ\sigma}_{ij} =\sum_{m{\bf k}}w_{m\bf k}f_{n\bf k} N^{IJ\sigma}_{\psi_{m{\bf k}}}
\langle \psi_{m{\bf k}}^\sigma |\phi^I_i \rangle \langle \phi^J_j |\psi_{m{\bf k}}^\sigma\rangle.
\label{Eq:RD}
\end{equation}
In case that there are no electrons participating the bond between atoms $I$ and $J$,
the renormalized density matrix of Eq.~\ref{Eq:RD} for the pair automatically reduces
to zero, thereby nullifying the inter-site effects. Note that $I$ and $J$ here also implicitly
include orbital indexes.

The bare Coulomb interaction between electrons belong to the pair
can be expressed by the electron repulsion integral~\cite{Agapito2015PRX},
\begin{equation}
V_\textrm{ERI} 
\equiv (ik|jl)\equiv\int d{\bf r}_1 d{\bf r}_2 
\frac{\phi^{I*}_i ({\bf r}_1)\phi^I_k({\bf r}_1)\phi^{J*}_j ({\bf r}_2)\phi^J_l ({\bf r}_2)}{|{\bf r}_1-{\bf r}_2|},
\label{Eq:ERI}
\end{equation}
where $i$ and $k$ are orbital indices belong to atom $I$ and $j$ and $l$ to atom $J$.
Using Eqs~(\ref{Eq:RD}) and (\ref{Eq:ERI}), the ACBN0-like energy expression 
($E^V_\textrm{ACBN0}$)
for the inter-site Hubbard interaction
can be written as 
\begin{eqnarray}
E^V_\textrm{ACBN0} &=& \frac{1}{4} \sum_{\{I,J\}} \sum_{ijkl} \sum_{\sigma,\sigma'} 
\left[
P^{II\sigma}_{ik}P^{JJ\sigma'}_{jl}-\delta_{\sigma \sigma'}P^{IJ\sigma}_{il}P^{JI\sigma'}_{jk}
\right]\nonumber \\
& &\times (ik|jl),
\label{Eq:ACBN0}
\end{eqnarray}
where the additional prefactor of 1/2 arises from a double counting of the same pairs.
Equating Eq.~(\ref{Eq:EVIJ}) to Eq.~(\ref{Eq:ACBN0}), 
then we can obtain a density functional form of $V^{IJ}$, 
\begin{equation}
V^{IJ}=\frac{1}{2}
\frac
{
\sum_{\sigma,\sigma'}\sum_{ijkl} 
[
P^{II\sigma}_{ik}P^{JJ\sigma'}_{jl}-\delta_{\sigma \sigma'}P^{IJ\sigma}_{il}P^{JI\sigma'}_{jk}
]
(ik|jl)
}
{
\sum_{\sigma,\sigma'}\sum_{ij} 
[
n^{II\sigma}_{ii} n^{JJ\sigma'}_{jj}-\delta_{\sigma \sigma'}n^{IJ\sigma}_{ij}n^{JI\sigma'}_{ji}
]
}.   	
\label{Eq:VF}
\end{equation}

An energy functional for $V$
can be constructed by subtracting a double counting term ($E_V^\textrm{dc}$) from $E_V$ in Eq.~(\ref{Eq:EVIJ}).
Following the discussion in Ref.~\onlinecite{Cococcioni2010JPC},
we use the fully localized limit and then 
$E_V^\textrm{dc}=\sum_{\{I,J\}}\sum_{i,j}\sum_{\sigma,\sigma'}
\frac{V^{IJ}}{2}n^{II\sigma}_{ii} n^{JJ\sigma'}_{jj}$.
With this, the final functional for the inter-site interaction of $V$ 
can be written as
\begin{equation}
E_\textrm{V} [\{ {\mathbf n} \}]=  
 - \frac{1}{2}\sum_{\{I,J\}} \sum_{\sigma}
 V^{IJ}[\{ {\mathbf n} \}] \textrm{Tr}[{\mathbf n}^{IJ\sigma}{\mathbf n}^{JI\sigma}],
\label{Eq:FF}
\end{equation}
where ${\mathbf n}^{IJ\sigma}$ is the matrix notation for the general occupation in Eq.~(\ref{Eq:GO}),
$\{ {\mathbf n} \}=\{{\mathbf n}^{II\sigma}$, ${\mathbf n}^{IJ\sigma}\}$
and $V^{IJ}[\{ {\mathbf n} \}]$ in Eq.~(\ref{Eq:VF}).
For the on-site repulsion, 
we used the ACBN0 functionals in Eqs (12) and (13) of Ref.~\onlinecite{Agapito2015PRX}
so that we complete a construction of the pseudohybrid-type functionals 
for the two essential Hubbard interactions.
As discussed before~\cite{Cococcioni2010JPC},
the minus sign in $E_\textrm{V} [\{ {\mathbf n} \}]$ 
highlights the role of the inter-site Hubbard interaction that localizes
electrons between atoms $I$ and $J$.
So, the equation~(\ref{Eq:FF}) implemented in this study can 
improve the description of covalent bonding 
or augmenting the overlocalization~\cite{Kulik2011JCP} caused by $U$
in case that the bonding between neighboring $d$- and $p$-orbitals plays important roles
for the various ground state properties of solids.

\section{Computational Methods\label{method}}

We implemented $E_\textrm{V} [\{ {\mathbf n} \}]$ in Eq.~(\ref{Eq:FF}), ACBN0 functionals 
and other related quantities in {\sc Quantum ESPRESSO} package~\cite{Giannozzi2009JPC}.
For the Kohn-Sham potential corresponding to Eq.~(\ref{Eq:FF}), 
we used Eq. (13) in Ref.~\onlinecite{Cococcioni2010JPC}.
To compute $V_\text{ERI}$ in Eq.~(\ref{Eq:ERI}), 
we used the PAO-3G minimal basis set as in Ref. \onlinecite{Agapito2015PRX}. 
With the aid of PyQuante package \cite{PyQuante}, the integrals were done quickly in an analytical way. 
For all calculations here, the cut-off for $d_{IJ}$ sets within the second-nearest neighbors.
We will discuss the effects of $d_{IJ}$ later.
The on-site interactions for $s$-orbitals were neglected for all materials considered here
while for inter-site interaction, $s$-orbitals were included.
Fully converged $U$ and $V$ were obtained when 
the difference in energy between two consecutive self-consistent steps
is less than 10$^{\text{-8}}$ Ry. 
We used the GBRV ultrasoft pseudopotentials \cite{Garrity2014CMS}.
Regarding pseudopotential dependence of ACBN0 functionals~\cite{Rubio2017PRB},
we tested the norm-conserving pseudopotentials provided by PseudoDoJo project \cite{Setten2018CPC}
and will discuss its effects in Appendix B.
The kinetic energy cutoff was set to 160 Ry to fix the value for all materials. 
The Brillouin zone (BZ) integration was performed 
with a ${\Gamma}$-centered $\mathbf k$-point grid spacing of 0.2 \AA$^{\text{-1}}$.
The lattice structures are chosen from the experimentally available data for comparison to the results with other computational methods and otherwise are relaxed within a standard DFT scheme.
For low dimensional materials discussed in Sec. V, we used slightly modified setups for computations and presented detailed 
methods in corresponding sections.

\begin{figure}[b]
	\begin{center}
		\includegraphics[width=\columnwidth]{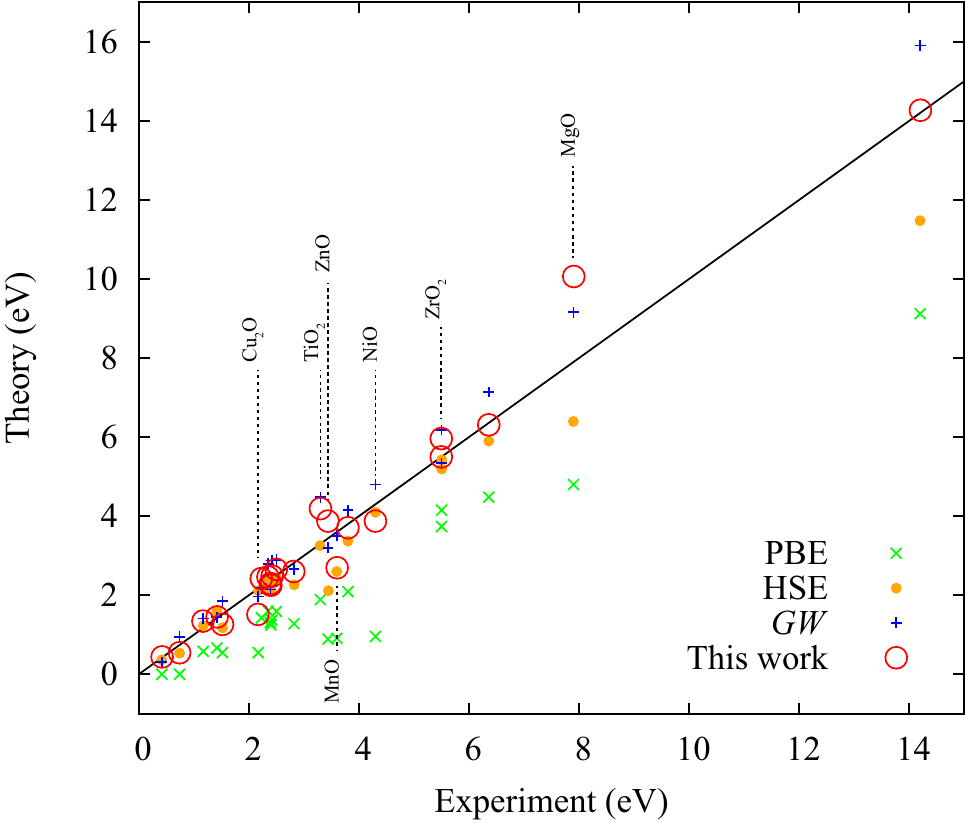}
	\end{center}
	\caption{Experimental versus theoretical band gaps in Table~\ref{table1}.
                 Metal oxides are marked 
                 and all other materials considered here are almost right on top of experimental 
                 values.}
	\label{Gap}
\end{figure}

\begin{table}[b]
\squeezetable
\caption{
Calculated band gaps (in eV). 
For comparisons, gaps from experiments and other methods are also shown. 
Structures denoted by the Strukturbericht designation are in parenthesis.
except monoclinic ZrO$_2$. 
Experimental data are 
from Refs.~\onlinecite{Lucero2012JPC,Tran2009PRL,Koller2011PRB,Agapito2015PRX} and references therein. 
}
\begin{ruledtabular} 
\begin{tabular}{ccccccc}
Solid   & GGA\footnote{GGA by Perdew-Burke-Ernzerhof (PBE)~\cite{PBE}.}  
& ACBN0 &   This Work
& HSE\footnote{
All data from HSE06 results in Ref. \cite{Garza2016JPCL} except 
ZrO$_2$ from HSE06 in Ref. \cite{Yuan2018JPC}, 
TiO$_2$ from HSE03 in Ref. \cite{Nakai2006JCCJ}, 
ZnO from HSE03 in Ref. \cite{Fuchs2007PRB} and 
Cu$_2$O from HSE06 in Ref. \cite{Heinemann2013PRB}
} 
& $GW$\footnote{
All data from self-consistent $GW$ (sc$GW$) calculation results in Ref.~\cite{Shishkin2007PRL} 
except GaP, InP, AlAs from sc$GW$ in Ref.~\cite{Remediakis1999PRB},
ZnSe from a partially self-consistent $GW$($GW_0$) in Ref.~\cite{Hinuma2014PRB}, 
BP from $GW_0$ in Ref.~\cite{Jiang2016PRB},
ZrO$_2$ from sc$GW$ in Ref.~\cite{Jiang2010PRB}, 
TiO$_2$ from sc$GW$ in Ref.~\cite{Lany2013PRB}, 
and Cu$_2$O from sc$GW$ in Ref.~\cite{Bruneval2006PRL}.
}
&  Expt. \\
\hline
C (A4)       & 4.15 &  4.17 &  5.50 &  5.43  &  6.18  &  5.50 \\        
Si (A4)      & 0.58 &  0.52 &  1.36 &  1.21  &  1.41  &  1.17 \\
Ge (A4)      & 0.00 &  0.00 &  0.61 &  0.80  &  0.95  &  0.74 \\
BP  (B3)     & 1.25 &  1.24 &  2.27 &  2.13  &  2.20  &  2.40 \\
AlP (B3)     & 1.59 &  2.00 &  2.66 &  2.42  &  2.90  &  2.50 \\
GaP (B3)     & 1.60 &  1.74 &  2.47 &  2.39  &  2.80  &  2.35 \\
InP (B3)     & 0.67 &  0.94 &  1.46 &  1.77  &  1.44  &  1.42 \\
AlAs (B3)    & 1.43 &  1.75 &  2.43 &  2.13  &  2.18  &  2.23 \\
GaAs (B3)    & 0.55 &  0.68 &  1.28 &  1.11  &  1.85  &  1.52 \\ 
InAs (B3)    & 0.00 &  0.00 &  0.46 &  0.57  &  0.31  &  0.42 \\
SiC  (B3)    & 1.37 &  1.74 &  2.49 &  2.32  &  2.88  &  2.42 \\
BN   (B3)    & 4.48 &  5.14 &  6.31 &  5.91  &  7.14  &  6.36 \\
ZnS  (B3)    & 2.09 &  3.43 &  3.71 &  3.44  &  4.15  &  3.80 \\
ZnSe (B3)    & 1.28 &  2.32 &  2.60 &  2.38  &  2.66  &  2.82 \\
ZnTe (B3)    & 1.31 &  1.99 &  2.30 &  2.34  &  2.15  &  2.39 \\
LiF  (B1)    & 9.12 & 13.74 & 14.26 & 13.28  & 15.90  & 14.20 \\
MgO  (B1)    & 4.80 &  8.84 & 10.06 &  6.59  &  9.16  &  7.90 \\
ZrO$_2$\footnote{Monoclinic structure} 
             & 3.74 &  5.10 &  5.97 &  5.20  &  5.34  &  5.50 \\  
TiO$_2$ (C4) & 1.89 &  3.02 &  4.18 &  3.25  &  4.48  &  3.30 \\ 
MnO (B1)     & 0.91 &  2.56 &  2.73 &  4.77  &  3.50  &  3.60 \\
NiO (B1)     & 0.96 &  3.70 &  3.90 &  4.09  &  4.80  &  4.30 \\ 
ZnO (B4)     & 0.89 &  3.62 &  3.88 &  2.11  &  3.80  &  3.44 \\
Cu$_2$O (C3) & 0.55 &  1.28 &  1.52 &  2.02  &  1.97  &  2.17 \\ 
\hline
MARE (\%)    &52.71 & 30.26 & 10.64  & 11.83  & 13.62   & \\
MRE (\%)     &$-$52.71 &$-$28.78 &0.47  &$-$2.61  & $7.76$ & 
\end{tabular}
\end{ruledtabular}
\label{table1}
\end{table}

\section{Energy gaps of three dimensional solids\label{3D}}

We first tested our method for selected bulk solids with diverse characteristics. 
Table~\ref{table1} and Fig.~\ref{Gap} summarize the calculated band gaps of 23 solids. 
We also listed the results from other calculations and measurements. 
We select solids from group IV, group III-V semiconductors, 
ionic insulators, metal chalcogenides and metal oxides.
Our calculated band gaps are in excellent agreement with experiments and are as accurate as those 
from HSE and $GW$ methods as shown in Table~\ref{table1}.
Mean absolute relative error (MARE) 
with respect to the experimental data indicates that
our method, HSE and $GW$ methods are closer to experiments than PBE and ACBN0. 
Mean relative error (MRE) shows that  
PBE and ACBN0 underestimate the gaps (minus sign) 
while $GW$ method overestimates them.
Hereafter, we mainly focus on the calculated gap values and, for future references,
the band structures of all solids are displayed in Supplementary Information~\cite{suppl}.

\begin{table}[b]
\caption{Calculated $U$ and $V$ between $s$- and $p$-orbitals of the first nearest 
neighbors of Si and GaAs (in eV). 
Here we compare our results with those based on the 
linear response theory (LRT).
For GaAs, the first (second) $U_p$ for on-site Hubbard interactions on Ga (As) $p$-orbital.
$V_{sp}$ ($V_{ps}$) corresponds to inter-site terms between Ga $s(p)$-orbital 
and As $p(s)$-orbital.
$V_{pp}$ denotes the inter-site interaction
between Ga $p$-orbital and $As$ $p$-orbital.
}
\begin{ruledtabular} 
\begin{tabular}{cccccccc}
        &   & $U_{p}$ & $V_{ss}$ & $V_{sp}$ & $V_{ps}$ & $V_{pp}$ \\
   \hline
 Si  & This work   & 3.50 & 0.90 & 0.72 & 0.72 & 1.85 \\
   & LRT\footnote{Reference \onlinecite{Cococcioni2010JPC}}  & 2.82 & 1.40 & 1.36 & 1.36 & 1.34 \\
 GaAs  & This work & 0.37, 1.88 & 0.91 & 1.26 & 0.80 & 1.75 \\
   & LRT$^\textrm{a}$  & 3.14, 4.24 & 1.75 & 1.76 & 1.68 & 1.72
\end{tabular}
\end{ruledtabular}
\label{U_Si_GaAs}
\end{table}

For the group IV semiconductors, 
the effect of $U$ on the band gaps is 
almost negligible as shown in Table~\ref{table1} (see ACBN0 column)
while the inter-site Hubbard terms improve
the band gaps dramatically as was also discussed 
in a previous study using the linear response 
theory~\cite{Cococcioni2010JPC}. 

For the group III-V semiconductors, 
both $U$ and $V$ affect their electronic structures 
because of their mixed covalent and ionic bonding characters. 
Therefore, ACBN0 improves the PBE gaps and the inter-site terms increases these further
to match experiment values.
Details of computations such as 
self-consistent $U$ and $V$ for Si and GaAs
compared with Ref.~\onlinecite{Cococcioni2010JPC}
are discussed below.
We note that PBE and ACBN0 incorrectly describe Ge and InAs as a metal
and a topological insulator, respectively,
while our method confirms them as semiconductors 
like HSE and $GW$ results.

We present the calculated $U$ and $V$ values for Si and GaAs 
to compare with the values obtained by the linear-response approach~\cite{Cococcioni2010JPC}. 
The Table~\ref{U_Si_GaAs} shows the calculated results. 
Our on-site $U$ value for $p$-orbital of Si is larger than that reported in a previous 
study~\cite{Cococcioni2010JPC}. 
All the other inter-site terms except one between $p$-orbitals
are smaller than those from the linear-response approach~\cite{Cococcioni2010JPC}.
We note here that the on-site term for Si has no effect on the band gap at all. 
Despite of the differences in the  Hubbard parameters, 
the calculated band gap of Si is in a good agreement 
with the previous linear response theory work~\cite{Cococcioni2010JPC}
and experiment value. 
In case of GaAs, our on-site interactions for $p$-orbitals of Ga and As, respectively, 
are all smaller than the previous results~\cite{Cococcioni2010JPC}. 
Like Si case, our inter-site values are also smaller than the previous results
except one between $p$-orbitals.
Nevertheless, our computed band gap of 1.28 eV for GaAs
is larger than the value of 0.90 eV reported in Ref.~\onlinecite{Cococcioni2010JPC}
and is close to the experimental gap of 1.52 eV (See Table~\ref{table1}).

In case of ionic compound LiF, 
the on-site $U$ improves the PBE band gap significantly 
because of its strong local Coulomb repulsion. 
Nonetheless, the inter-site $V$ still increases the ACBN0 gap further
to match with an experimental value.

A similar trend is also found in metal monochalcogenides
(here Zn compounds only).
For these compounds, the $U$ and $V$ functionals play similar roles 
as they do for LiF so that our results with $U$ and $V$
are quite closer to experiment values than those with $U$ only.
We note that the calculated gaps depends on the choice of pseudopotential of Zn
while there is no such dependence in cases of IV and III-V semiconductors.
We will discuss this further for cases of metal oxides below.

Regarding metal oxides, 
our results agree with the calculations by other advanced methods. 
For TiO$_2$, MnO, NiO and ZnO
in Table~\ref{table1},
our ACBN0 results already improves PBE gaps significantly, similar with
previous studies~\cite{Agapito2015PRX,Gopal2017JPCM,Rubio2017PRB} that calculated
the detailed electronic structures with ACBN0.
Our gaps are slightly larger than the values in other works~\cite{Agapito2015PRX,Gopal2017JPCM,Rubio2017PRB}. 
This discrepancies originate
from the different self-consistent $U$ values. 
With the inter-site $V$ included, the changes in the on-site $U$ 
lead to increase in the ACBN0 gaps of TiO$_2$, MnO, ZnO and NiO as shown in Table~\ref{U_Oxide}.
We note that the gaps of metal oxides 
depend on the choice of pseudopotentials.
With the potentials from the PseudoDoJo project~\cite{Setten2018CPC},
we achieve a better agreement (Table~\ref{Table_A2} in Appendix). 
Because effects of on-site and inter-site interactions depend
on degree of localization or cut-off in projector for localized orbital~\cite{Rubio2017PRB},
it is important to select or generate pseudopotentials 
with care to obtain accurate results \cite{Kulik2008JCP}
or to develop a computational method for the on- and inter-site Hubbard interactions
that do not depend on projectors. 

For Cu$_\text{2}$O and Zr$_2$O, our results are comparable to those from 
HSE and $GW$ calculations. 
We note here that for Cu$_2$O 
the calculated energetic position of fully filled $d$-orbitals
is lower than that of HSE value~\cite{Heinemann2013PRB} by $\sim0.9$ eV [Fig. S7 in SI]. 
Since the degree of screening depends on occupancy of orbitals in the ACBN0 formalism, 
the weak screening for the fully filled $d$-orbitals like cuprite seems to be inevitable. 
Thus, it needs to improve the way to treat 
the completely filled $d$-orbitals with $U$ and $V$ within this formalism.
Regardless of its limitation, we found that our computed gaps with $V$ are considerably 
improved if compared with those from ACBN0 and mBJLDA.
Considering limits in mBJLDA to obtain gaps for these compounds~\cite{Koller2011PRB},
our method could be a good alternative tool for studying zirconia and cuprite.

\begin{table}[b]
\caption{Self-consistent on-site energies of $3d$ orbitals of transition metal ($U_d$)
and $2p$ orbitals of oxygen ($U_p$) and direct band gaps ($E_g$) of selected transition metal oxides
within the ACBN0 formalism (in eV).
We also list the self-consistent on-site and 
inter-site Hubbard interactions ($V_{dp}$) between the metal $d$ 
and oxide $p$ orbitals, and $E_g$ using the current method. 
}
\begin{ruledtabular} 
\begin{tabular}{cccccc}
Solids  &                            & $U_d$        & $U_p$        &$V_{dp}$ & $E_g$  \\
\hline                                                       
TiO$_2$ & This work\footnote{DFT+$U$}                                              &  0.27 &  8.49 &         & 3.02 \\
               & ACBN0\footnote{Reference~\onlinecite{Agapito2015PRX}} &  0.15 &  7.34 &         & 2.83 \\
              & ACBN0\footnote{Reference~\onlinecite{Rubio2017PRB}}   &  0.96  & 10.18 &        & 3.21 \\
              & This work\footnote{DFT+$U$+$V$}                                      &  0.37 &  8.21 & 2.94 & 4.18 \\
\hline                                                       
MnO     & This work$^\textrm{a}$     &  5.11 &  2.99  &        &  3.05 \\
             & ACBN0$^\textrm{b}$        &  4.67  &  2.68 &       &   2.83 \\
             & ACBN0$^\textrm{c}$        &  4.68  &  5.18  &       &  2.65 \\
             & This work$^\textrm{d}$     &  5.31 &  2.94 & 2.72 & 3.60\\
\hline                                                       
NiO     & This work$^\textrm{a}$       &  8.22  &  2.83 &         & 4.66 \\    
            & ACBN0$^\textrm{b}$           &  7.63  &  3.00 &        & 4.29 \\
           & ACBN0$^\textrm{c}$           &  6.93  &  2.68 &         & 4.14 \\ 
            & This work$^\textrm{d}$       &  7.77  &  2.37 & 2.93 & 5.13 \\    
\hline                                                       
ZnO     & This work$^\textrm{a}$      & 15.06  &  7.30 &         &  3.62 \\
            & ACBN0$^\textrm{b}$          & 12.80  &  5.29 &         & 2.91  \\
            & ACBN0$^\textrm{c}$          & 13.30  &  5.95 &         & 3.04  \\
             & This work$^\textrm{d}$      & 14.96  &  7.07 & 3.01 & 3.88  \\
\end{tabular}
\end{ruledtabular}
\label{U_Oxide}
\end{table}

To compare with the previous ACBN0 studies on metal oxides~\cite{Agapito2015PRX, Rubio2017PRB}, 
we consider the on-site Hubbard interaction 
of $U_d$ for $d$ electrons of metals and 
$U_p$ for $p$ electrons of oxygen in TiO$_2$, MnO, NiO and ZnO, respectively. 
We also provide those values and the first-nearest neighbor 
inter-site Hubbard interaction terms (the $d$-$p$ interactions) 
calculated with our method. 
The results are summarized in Table~\ref{U_Oxide}. 
We note that except TiO$_2$ our on-site repulsions for 
$d$-orbitals of metals are rather larger than the values from the previous works~\cite{Agapito2015PRX, Rubio2017PRB}
while the repulsions for oxide $p$ orbitals are similar with 
the previous studies. 
Since the size of on-site repulsion of $d$-orbital is almost proportional to the size of gap, 
our larger estimations of $U_d$ result in relatively larger direct energy gaps for 
MnO, NiO and ZnO, respectively if compared with the previous studies~\cite{Agapito2015PRX, Rubio2017PRB}.
With including the inter-site $V_{dp}$, the gaps increase further
because of reduction in energetic position of conduction band maximum
at $\Gamma$-point (See Figs. 6S in SI). 
As mentioned in Ref.~\onlinecite{Rubio2017PRB}, 
the discrepancies of $U_d$, $U_p$ and $E_g$ may be
attributed to the way to calculate Coulomb integrals 
and the treatment of the localized orbitals. 
As shown in Table~\ref{Table_A2}, our gap values also change
according to the different choice of pseudopotentials. 
Therefore, a further study on this problem is required to obtain
better description of electronic structures of metal oxides. 

\section{Energy gaps in low-dimensional materials\label{2D}}

Now, we consider low-dimensional systems
where the screening of Coulomb interaction varies rapidly.  
The $GW$ approximation calculates quasiparticle gaps
quite accurately but its convergence is very slow with respect 
to the $\mathbf{k}$-points density and 
other parameters~\cite{Qiu2016PRB,Rasmussen2016PRB}.
The hybrid functional methods do not suffer such an issue but
they produce unreliable gap values with structural or dimensional 
variations~\cite{Jain2011PRL}.
The mBJLDA, another low-cost alternative for bulk solids,
also suffer a similar problem as hybrid functionals~\cite{Ji2014NatComm}.
However, with our method, self-consistently computed occupations
of atoms at boundary and bulk reflect  the screening of Coulomb interaction through Eqs.~(\ref{Eq:RON}), (\ref{Eq:RD}) and (\ref{Eq:VF}).
Hence, we expect that the current method may overcome the aforementioned difficulties
for low dimensional materials. 

\begin{figure}[t]
	\begin{center}
		\includegraphics[width=0.8\columnwidth]{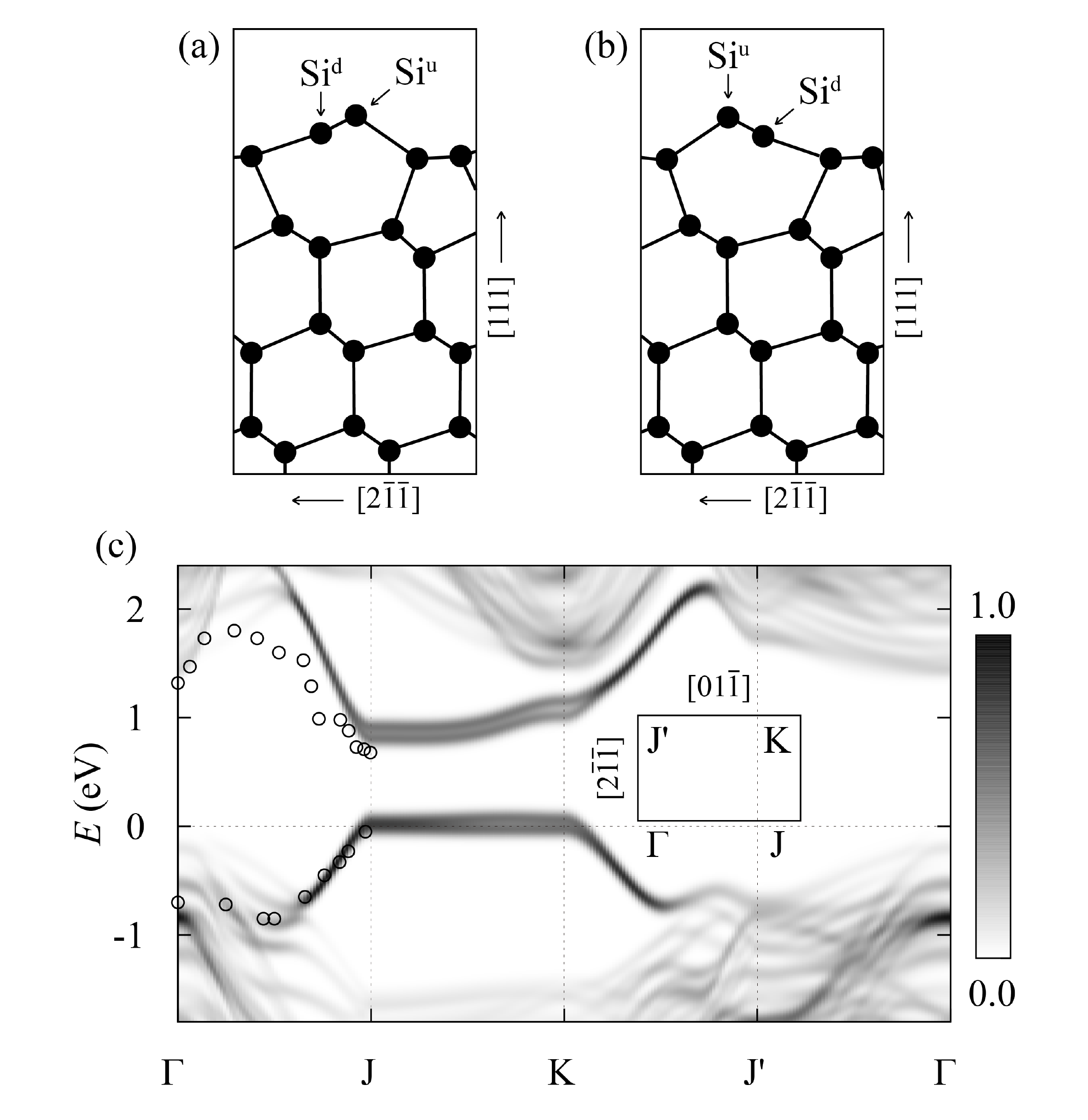}
	\end{center}
	\caption{
	The cross-sectional view of optimized atomic structures for two energetically degenerate
	buckled structures are shown in (a) and (b). 
          Si atom (filled circles) relaxed down to (up away from) the surface is denoted by Si$^\textrm{d}$ and
          Si$^\textrm{u}$, respectively.
	(c) Averaged surface band structures projected to
         the first four layers of Si(111)-($2\times 1$) where the scale on the right side denotes 
         local density of states in an arbitrary unit. 
         The energetic position of bulk valence band maximum at $\Gamma$ is set to zero.
         Black open circles are experimental data 
         from direct~\cite{Uhrberg1982PRL}
         and inverse~\cite{Perfetti1987PRB} photoemission spectroscopies experiments.
         The inset describes the BZ. 
	}
	\label{Si}
\end{figure}

To test the new method, we first calculated the electronic structures of Si(111)-($2\times 1$) surface.
Because of a unique surface reconstruction resulting 
in a quasi-one-dimensional $\pi$-bonded chain of Si $p_z$-orbitals~\cite{Pandey1981prl}
and a large difference between the  screenings on surface and in bulk, 
it is a good test bed for a method to compute surface 
and bulk gap simultaneously~\cite{Rohlfing1999PRL,Jain2011PRL}.
A 24-layer slab with $\sim$15 \AA ~vacuum 
was optimized with GGA until the residual forces on atoms are less than $10^{-4}$ Ry/\AA. 
The kinetic energy cutoff is set to 80 Ry and 
$d_{IJ}$ to the nearest neighbors.
The surface has two degenerate and coexisting 
reconstructions~\cite{rohlfing2012prb,violante2014ss} as shown in Figs.~\ref{Si} (a) and (b)
so that we compute the averaged surface band structures to compare with experiments.
As shown in Fig.~\ref{Si} (c), the calculated averaged surface gap is
0.83 eV, agreeing well with the  experimental value of 0.75 eV~\cite{Uhrberg1982PRL, Perfetti1987PRB}, 
together with an accurate bulk gap. 
We note that the converged Hubbard parameters 
of Si atoms change spatially, reflecting the local variation of screening 
such that the calculated $U$ and $V$ are confirmed 
to gradually increase from inside to the surface (not shown here).

\begin{figure}[t]
	\begin{center}
		\includegraphics[width=0.8\columnwidth]{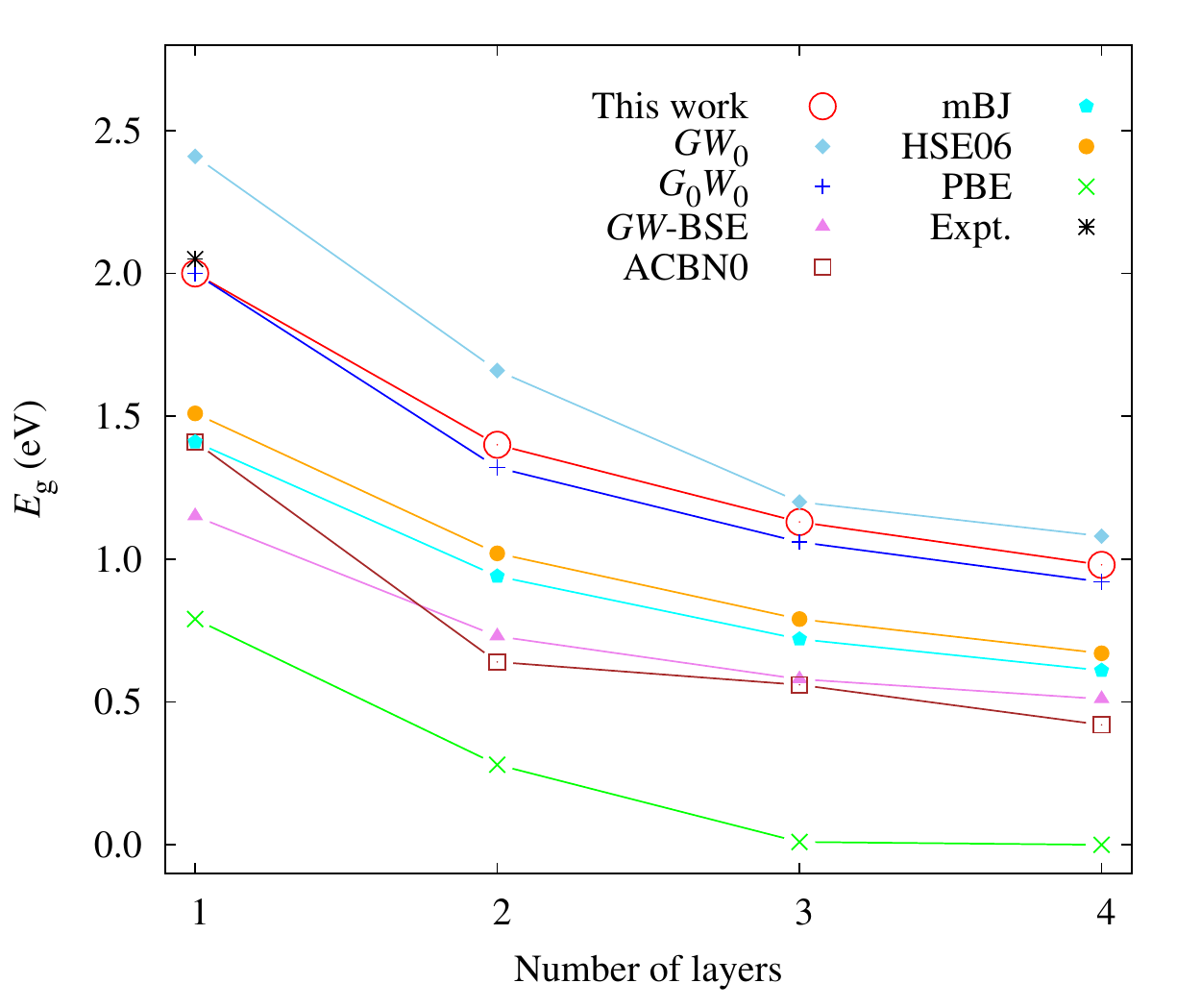}
	\end{center}
	\caption{Band gaps of black phosphorus as a function of a number of layers. 
        We also list other gaps from PBE, 
        HSE06~\cite{Ji2014NatComm}, mBJLDA~\cite{Ji2014NatComm}, $GW_{0}$~\cite{Wang2015PRB},
        $G_0 W_0$~\cite{Rudenko2015PRB} and $GW$-BSE~\cite{Rudenko2015PRB}. 
        A experimental band gap~\cite{Liang2014nanolett} is denoted with a black cross.}
	\label{BP}
\end{figure}

Next, few-layer black phosphorus (BP) 
was chosen to test our method [Fig.~\ref{BP}].
We used fully relaxed crystal structures 
using the rev-vdW-DF2 functional~\cite{Hamada2014PRB,Kim2017PRB}.
All inter-site interactions between
valence $s$ and $p$ electrons of P atom within the plane are considered.
Fig.~\ref{BP} shows the calculated 
band gaps in terms of the number of layers, together
with other calculations and experiment. 
It is noticeable that, without including $V$, 
all ACBN0 gaps are quite smaller than $GW$ gaps, 
and that 
HSE~\cite{Ji2014NatComm}, mBJLDA~\cite{Ji2014NatComm} 
and ACBN0 produce the gaps close to the optical gaps by $GW$-BSE method~\cite{Rudenko2015PRB}.
Considering qualitative difference in shape of optical spectrum 
between $GW$-BSE and HSE or other hybrid functionals~\cite{Jain2011PRL,Li2014PRB},
we conclude that they underestimate
band gaps.

As shown in Fig.~\ref{BP}, our results are consistent with $GW$ results~\cite{Wang2015PRB,Rudenko2015PRB}
and an available experiment~\cite{Liang2014nanolett}.
We note that the computed band gaps of pure 2D materials such as single layer BP 
have a slight dependence
on the range of inter-site Hubbard interaction (will be discussed in the next Section), 
reflecting complex nature of screening in low dimensional materials~\cite{Qiu2016PRB}.
Considering a large number of atoms
in typical nanostructures, 
our new method will have a merit over the other computational methods
that requires quite expensive computational resources.

\section{Discussion and Conclusion\label{discussion}}

\begin{figure}[b]
	\begin{center}
		\includegraphics[width=1.0\columnwidth]{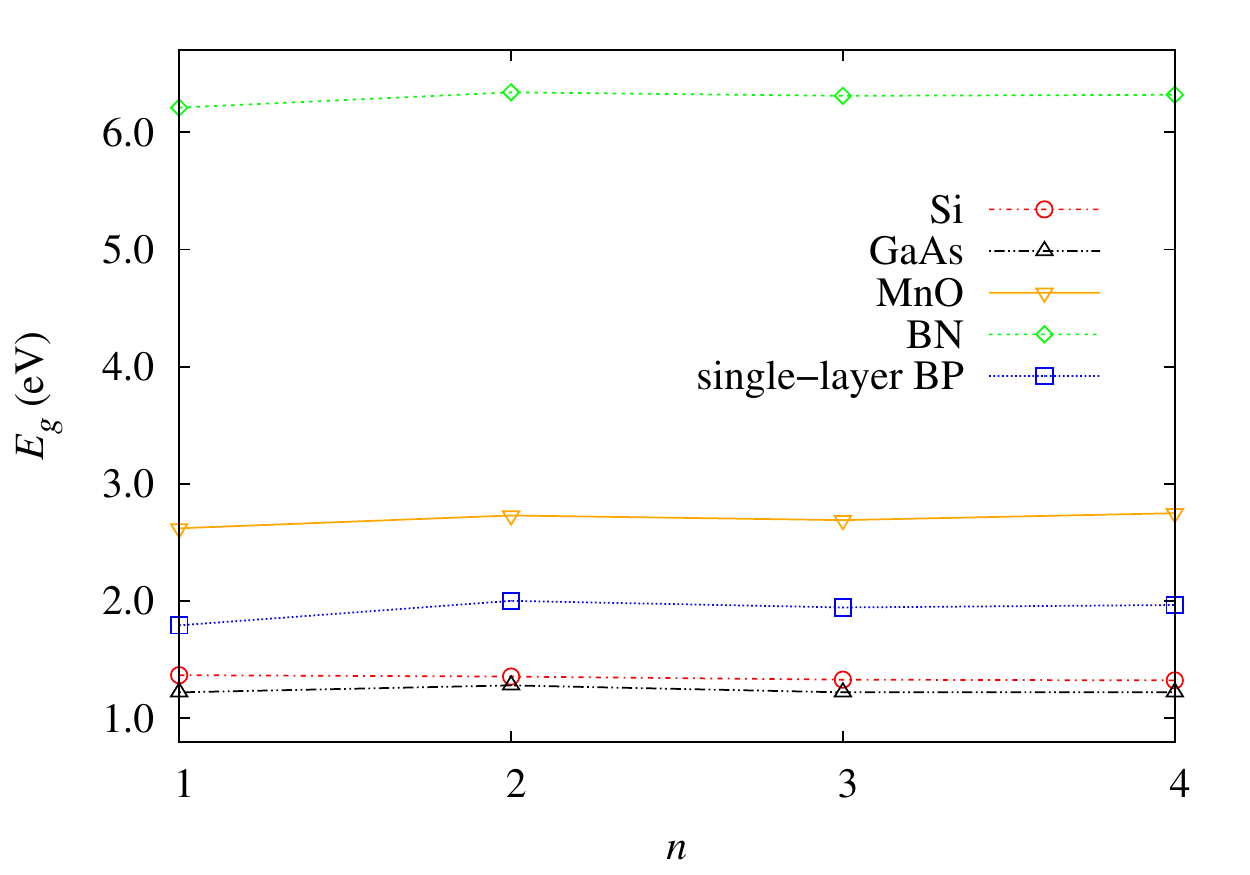}
	\end{center}
	\caption{Energy gap variations of Si, GaAs, MnO, BN and a single layer BP 
	as a function of the distance between the pair atoms for the inter-site Hubbard interaction.
	The abscissa denotes $n$-th nearest neighbor and the ordinate shows the gap with including $V$'s up to the $n$-th neighbor. }
	\label{dij}
\end{figure}

The only empirical parameter in the present formalism is $d_{IJ}$
that determines the range of a pair for the inter-site $V$
in Eqs.~(\ref{Eq:EVIJ}) and~(\ref{Eq:FF}). For the three dimensional (3D) solids
studied here, we find that the nearest neighbor inter-site interaction is enough to obtain
the converged band gap values. As shown in Fig.~\ref{dij}, the energy band 
gap of 3D Si, GaAs and MnO crystals show a negligible variation as increasing $d_{IJ}$
while the zincblende structure BN shows a converged gap after including the second nearest neighbors.
Unlike the most 3D cases, the energy gaps for some low dimensional systems show 
rather larger variations as a function of $d_{IJ}$ than 3D materials do. 
For the case of the reconstructed
Si(111)-($2\times1$) surface discussed in the previous section, 
we don't need to include an inter-site $V$ for the next nearest neighboring Si atoms. 
However, for a single layer BP shown in Fig.~\ref{dij}, 
the gap with $V$ for the nearest neighbors is smaller (about 10\%) than one with $V$ for the next ones.  
Beyond this, the gap varies a little so that at least two sets of $V$ with different $d_{IJ}$
are required to obtain a reasonably converged band gap of the single layered BP.
We note here that the increase in computational time with a longer $d_{IJ}$ is amount
to the increase in the DFT-GGA computation for a corresponding larger supercell case.

\begin{table}[t]
\caption{Calculated magnetic moments in $\mu_B$ of antiferromangetic MnO and NiO and their comparisons with other works
and experiments. Here the moments are projected values for one spin orientation.
}
\begin{ruledtabular} 
\begin{tabular}{ccc}
                 &   MnO & NiO \\
\hline
   
This work ($+V$)  & 4.69 & 1.78  \\
HSE~\cite{Rodl2009PRB} & 4.5 & 1.5 \\
ACBN0-PBE~\cite{Gopal2017JPCM} &4.79 & 1.83 \\
Experiments &4.58~\cite{Cheetham1983PRB}, 4.79~\cite{Fender1968JCP} & 
  1.77~\cite{Fender1968JCP},  1.90~\cite{Roth1958PR,Cheetham1983PRB}
\end{tabular}
\end{ruledtabular}
\label{MM}
\end{table}

The role of self-consistent inter-site Hubbard interactions 
on magnetic moments is also an interesting issue. 
For this, we calculated magnetic moments of antiferromagnetic MnO and NiO. 
As shown in Table~\ref{MM}, the calculated magnetic moments 
are slightly reduced from the values using ACBN0-PBE method~\cite{Gopal2017JPCM} 
where the on-site $U$ enhance the localization of electrons at atomic sites. 
On the other hand, the inter-site $V$ reduces the on-site localization and shifts
electrons to the bonding sites.
Therefore, the competition between $U$ and $V$ gives rise 
to the reduced magnetic moments compared with ACBN0 method 
and the calculated moments are 
in excellent agreement with experiments~\cite{Cheetham1983PRB, Fender1968JCP,Roth1958PR}.

In conclusion, we report a new {\it ab initio} 
method for electronic structures of solids 
employing a pseudohybrid density functional for extended Hubbard Coulomb interactions.
We demonstrate that the new method significantly 
improves the original ACBN0 functional in obtaining band gaps
of bulk  and low dimensional materials.
Its self-consistent calculation can be done with a computational time comparable to DFT-GGA.
With further validations with other methods~\cite{Huang2020PRB, Timrov2020arXiv} and improvements of the current method 
such as the noncollinear spin and forces~\cite{Rubio2017PRB},
our new method could fulfill requirements~\cite{Curtarolo2013NatMat} for 
first-principles simulations suitable 
for massive database-driven materials research with an improved accuracy.

\section*{Acknowledgments}

We thank B.-H. Kim, H.-J. Kim, S.-H. Kang, S. Kim, S. Y. Park, S.-H. Jhi, H. J. Choi and M. J. Han for fruitful discussions.
Y.-W.S was supported by NRF of Korea (Grant No. 2017R1A5A1014862, SRC program: vdWMRC center) and KIAS individual grant (CG031509).
Computations were supported by the CAC of KIAS. This work was also supported by the National Supercomputing Center with supercomputing resources including technical support (KSC-2018-C2-0017).

\appendix
\setcounter{table}{0}
\renewcommand{\thetable}{A\Roman{table}}%

\section{Effects of cross exchange interactions between orbitals}

\begin{table}[b]
\squeezetable
\caption{
Calculated band gaps (in eV) with and without the cross charge exchange
in Eq.~(\ref{EqS:KIJ}). `$+V$' column summarizes the band gap with $V^{IJ}$ and without $K^{IJ}$ 
while `$+V_\textrm{eff}$'column with $V_\text{eff}^{IJ}=V^{IJ}-K^{IJ}$. 
Experimental data for energy gaps 
are from Refs.~\cite{Lucero2012JPC,Tran2009PRL,Koller2011PRB,Tezuka1994JPSJ,Sawatzky1984PRL,Agapito2015PRX}. 
}
\begin{ruledtabular} 
\begin{tabular}{cccccc}
Solid    & ACBN0 &  This Work  & This Work 
& $GW$\footnote{
All data from self-consistent $GW$ (sc$GW$) calculation results in Ref.~\cite{Shishkin2007PRL} 
except GaP, InP, AlAs from sc$GW$ in Ref.~\cite{Remediakis1999PRB},
ZnSe from $G_0 W_0$ in Ref.~\cite{Hinuma2014PRB}, 
ZrO$_2$ from sc$GW$ in Ref.~\cite{Jiang2010PRB}, 
TiO$_2$ from sc$GW$ in Ref.~\cite{Lany2013PRB}, 
and Cu$_2$O from sc$GW$ in Ref.~\cite{Bruneval2006PRL}.
}
&  Expt. \\
     &      &  ($+V$)   &  ($+V_\textrm{eff}$) &   &\\
\hline
C (A4)       &  4.17 &  5.50 &   5.36  &   6.18  &  5.50 \\        
Si (A4)      &  0.52 &  1.36 &   1.24  &   1.41  &  1.17 \\
Ge (A4)      &  0.00 &  0.61 &   0.48  &   0.95  &  0.74 \\
BP  (B3)     &  1.24 &  2.27 &   2.14  &   2.20  &  2.40 \\
AlP (B3)     &  2.00 &  2.66 &   2.58  &   2.90  &  2.50 \\
GaP (B3)     &  1.74 &  2.47 &   2.39  &   2.80  &  2.35 \\
InP (B3)     &  0.94 &  1.46 &   1.41  &   1.44  &  1.42 \\
AlAs (B3)    &  1.75 &  2.43 &   2.35  &   2.18  &  2.23 \\
GaAs (B3)    &  0.68 &  1.28 &   1.21  &   1.85  &  1.52 \\ 
InAs (B3)    &  0.00 &  0.46 &   0.41  &   0.31  &  0.42 \\
SiC  (B3)    &  1.74 &  2.49 &   2.39  &   2.88  &  2.42 \\
BN   (B3)    &  5.14 &  6.31 &   6.17  &   7.14  &  6.36 \\
ZnS  (B3)    &  3.43 &  3.71 &   3.68  &   4.15  &  3.80 \\
ZnSe (B3)    &  2.32 &  2.60 &   2.57  &   2.66  &  2.82 \\
ZnTe (B3)    &  1.99 &  2.30 &   2.28  &   2.15  &  2.39 \\
LiF  (B1)    & 13.74 & 14.26 &  14.22  &  15.90  & 14.20 \\
MgO  (B1)    &  8.84 & 10.06 &   9.93  &   9.16  &  7.90 \\
ZrO$_2$      &  5.10 &  5.97 &   5.95  &   5.34  &  5.50 \\  
TiO$_2$ (C4) &  3.02 &  4.18 &   4.16  &   4.48  &  3.30 \\ 
MnO (B1)     &  2.56 &  2.73 &   2.71  &   3.50  &  3.60 \\
NiO (B1)     &  3.70 &  3.90 &   3.80  &   4.80  &  4.30 \\ 
ZnO (B4)     &  3.62 &  3.88 &   3.86  &   3.80  &  3.44 \\
Cu$_2$O (C3) &  1.28 &  1.52 &   1.50  &   1.97  &  2.17 \\ 
\hline                                                                             
MARE (\%)    & 30.26 & 10.64 &  10.81  & 13.62   & \\
MRE (\%)     & $-$28.78 &0.47 &$-$3.11  & 7.76 & 
\end{tabular}
\end{ruledtabular}
\label{Table_A1}
\end{table}

The mean-field (MF) expression of the electronic interaction energy in terms of atomic orbitals in its most general
form can be written as , 
\begin{equation}
E_\textrm{MF}=\frac{1}{2}\sum_{I,J,K,L}\sum_{ijkl} \sum_{\sigma\sigma'}E^{IJKL}_{ijkl,\sigma\sigma'},
\label{EqS:MF}
\end{equation}
where,
\begin{eqnarray}
E_{ijkl,\sigma\sigma'}^{IJKL}&=&  
\langle \phi^I_i \phi^J_j | V_{ee} | \phi^K_k \phi^L_l \rangle  \nonumber \\
& &\times
\left( n^{KI\sigma}_{ki} n^{LJ\sigma'}_{lj}
  -\delta_{\sigma \sigma'} n^{KJ\sigma}_{kj} n^{LI\sigma'}_{li} \right),
\label{EqS:HF}
\end{eqnarray}
where $n^{IJ\sigma}_{ij}$ is defined in Eq. (2). 

Considering $E_{ijkl,\sigma\sigma'}^{IJKL}$, there are many possible arrangements for $IJKL$
and $ijkl$~\cite{Cococcioni2010JPC}, respectively. 
Among them, here we consider the first three large contributions, 
$E_{ijij,\sigma\sigma'}^{IIII}$, $E_{ijij,\sigma\sigma'}^{IJIJ}$ and $E_{ijji,\sigma\sigma'}^{IJJI}$
where $I\neq J$.
The first and second terms were discussed in the main manuscript and corresponds to the on-site
and the inter-site Hubbard interactions, respectively. 
The last one is the cross charge exhanges between the neighboring atoms $I$ and $J$~\cite{Cococcioni2010JPC}.
The first case where all $IJKL$ are equals will lead to the well known Hubbard density functional for LDA+$U$ method
and if we use a rotationally invariant on-site interaction ($E^I_U$), 
the Eq.~(\ref{EqS:MF}) will become to be Dudarev $U$ functional~\cite{Dudarev1998PRB}.
The second term becomes the inter-site Hubbard interaction as discussed in the main manuscript.

\begin{table}[b]
\squeezetable
\caption{
Calculated band gaps (in eV) with two different sets of pseudopotentials.
We tested GBRV ultrasoft pseudopotentials \cite{Garrity2014CMS} 
and the norm-conserving pseudopotentials provided by PseudoDoJo project \cite{Setten2018CPC}.
Two `$V$' columns summarize the calculated band gaps with the current method.
Experimental data for energy gaps 
are from Refs.~\cite{Lucero2012JPC,Tran2009PRL,Koller2011PRB,Tezuka1994JPSJ,Sawatzky1984PRL,Agapito2015PRX}. 
}
\begin{ruledtabular} 
\begin{tabular}{cccccc}
Solid    
& ACBN0\footnote{PseudoDoJo pseudopotential} 
& ACBN0\footnote{GBRV ultrasoft pseudopotential} 
& $+V$ \footnote{PseudoDoJo pseudopotential} 
& $+V$ \footnote{GBRV ultrasoft pseudopotential} &  Expt. \\
\hline
C (A4)       &  4.21      &      4.17     &    5.54     &   5.50   &  5.50 \\           
Si (A4)      &  0.52      &      0.52     &    1.35     &   1.36   &  1.17 \\
Ge (A4)      &  0.00      &      0.00     &    0.60     &   0.61   &  0.74 \\
BP  (B3)     &  1.24      &      1.24     &    2.27     &   2.27   &  2.40 \\
AlP (B3)     &  1.99      &      2.00     &    2.66     &   2.66   &  2.50 \\
GaP (B3)     &  1.78      &      1.74     &    2.47     &   2.47   &  2.35 \\
InP (B3)     &  1.00      &      0.94     &    1.46     &   1.46   &  1.42 \\
AlAs (B3)    &  1.76      &      1.75     &    2.43     &   2.43   &  2.23 \\
GaAs (B3)    &  0.70      &      0.68     &    1.22     &   1.28   &  1.52 \\ 
InAs (B3)    &  0.00      &      0.00     &    0.44     &   0.46   &  0.42 \\
SiC  (B3)    &  1.73      &      1.74     &    2.49     &   2.49   &  2.42 \\
BN   (B3)    &  5.21      &      5.14     &    6.39     &   6.31   &  6.36 \\
ZnS  (B3)    &  4.30      &      3.43     &    4.77     &   3.71   &  3.80 \\
ZnSe (B3)    &  3.16      &      2.32     &    3.62     &   2.60   &  2.82 \\
ZnTe (B3)    &  2.73      &      1.99     &    3.28     &   2.30   &  2.39 \\
LiF  (B1)    & 14.88      &     13.74     &   16.01     &  14.26   & 14.20 \\
MgO  (B1)    &  8.89      &      8.84     &   10.16     &  10.06   &  7.90 \\
ZrO$_2$      &  5.32      &      5.10     &    6.19     &   5.97   &  5.50 \\  
TiO$_2$ (C4) &  2.89      &      3.02     &    4.08     &   4.18   &  3.30 \\ 
MnO (B1)     &  2.88      &      2.56     &    3.55     &   2.73   &  3.60 \\
NiO (B1)     &  4.30      &      3.70     &    4.82     &   3.90   &  4.30 \\ 
ZnO (B4)     &  4.38      &      3.62     &    4.91     &   3.88   &  3.44 \\
Cu$_2$O (C3) &  1.63      &      1.28     &    2.02     &   1.52   &  2.17 \\ 
\hline                                                                                
MARE (\%)    & 29.04      &     30.26     &   14.06     &  10.64   & \\
MRE (\%)     & $-$21.73      &     $-$28.78     &   9.51   &   0.47   & 
\end{tabular}
\end{ruledtabular}
\label{Table_A2}
\end{table}

Now, we consider the second and third ones together. 
If we use rotationally invariant forms for 
$\langle \phi^I_i \phi^J_j | V_{ee} | \phi^K_k \phi^L_l \rangle$ in Eq.~(\ref{EqS:HF}), 
we can rewrite the second interaction using 
$
\langle \phi^I_i \phi^J_j | V_{ee} | \phi^K_k \phi^L_l \rangle
= V^{IJ}\delta_{IK}\delta_{JL}\delta_{ik}\delta_{jl}
$
and 
$
N^{IJ}V^{IJ}=\sum_{i,j}
\langle \phi^I_i \phi^J_j | V_{ee} | \phi^I_i \phi^J_j \rangle
$
where $N^{IJ}$ is a number of degeneracy of angular momentum for atoms $I$ and $J$~\cite{Cococcioni2010JPC}.
Likewise, the third interactions, we can use
$
\langle \phi^I_i \phi^J_j | V_{ee} | \phi^K_k \phi^L_l \rangle
= K^{IJ}\delta_{IL}\delta_{JK}\delta_{il}\delta_{jk}
$
and 
$
N^{IJ}K^{IJ}=\sum_{i,j}
\langle \phi^I_i \phi^J_j | V_{ee} | \phi^J_j \phi^I_i \rangle.
$
With these considerations, the MF energy in Eq.~(\ref{EqS:MF}) can be written as 
\begin{equation}
E_\textrm{Hub}=\frac{1}{2}\left[\sum_I E^I_U+\sum_{\{I,J\}}E^{IJ}_V
+\sum_{\{I,J\}}E^{IJ}_K\right],
\end{equation}
where the $\{I,J\}$ indicates the summation 
for a pair of atoms $I$ and $J$ within a given cut-off of $d_{IJ}$.

Using the matrix notations of ${\mathbf n}^{IJ}=\sum_\sigma{\mathbf n}^{IJ\sigma}$ 
and $n^I=\sum_\sigma\sum_i n^{II\sigma}_{ii}$
for the general occupation in Eq. (2),
\begin{eqnarray}
E_V^{IJ}+E_K^{IJ}
&=& V^{IJ}\left[n^I n^J -\sum_\sigma\textrm{Tr}[{\mathbf n}^{IJ\sigma}{\mathbf n}^{JI\sigma}]\right] \nonumber\\
&+& K^{IJ}\left[\textrm{Tr}[{\mathbf n}^{IJ}{\mathbf n}^{JI}] -\sum_\sigma n^{I\sigma} n^{J\sigma}]\right]
\label{EqS:EVEK}
\end{eqnarray}
We assume the fully localized limit for double counting as was also discussed in a previous work~\cite{Cococcioni2010JPC} so that
the final expression for Hubbard pairwise energy is given by, 
\begin{eqnarray}
&  & E_V^{IJ}+E_K^{IJ}-E_{dc} \nonumber \\
&=& -V^{IJ}\sum_\sigma\textrm{Tr}[{\mathbf n}^{IJ\sigma}{\mathbf n}^{JI\sigma}] 
+K^{IJ}\textrm{Tr}[{\mathbf n}^{IJ}{\mathbf n}^{JI}] \nonumber \\
&\simeq& -(V^{IJ}-K^{IJ}) \sum_\sigma\textrm{Tr}[{\mathbf n}^{IJ\sigma}{\mathbf n}^{JI\sigma}],
\label{EqS:FHub}
\end{eqnarray}
where we neglect
$K^{IJ}\sum_{\sigma\neq\sigma'}\textrm{Tr}[{\mathbf n}^{IJ\sigma}{\mathbf n}^{JI\sigma'}]$
thanks to $V^{IJ}-K^{IJ}\gg K^{IJ}$.

If we compare Eq.~(\ref{EqS:FHub}) with the inter-site Hubbard functional shown in Eq. (9) in the main manuscript, 
we immediately notice that a replacement of $V^{IJ}$ by $V^{IJ}_\textrm{eff}=V^{IJ}-K^{IJ}$ is enough
for including the effects of cross charge exchange between the pair of atoms $I$ and $J$.

Now using Eqs. (4)-(6) in the main manuscript, 
a pseudohybrid or ACBN0-like functional expression for $K^{IJ}$
can be obtained in a straightforward way and the final expression can be written as 
\begin{equation}
K^{IJ}=\frac{1}{2}\frac{
                          \sum_{\sigma,\sigma'}\sum_{ijkl} [\delta_{\sigma \sigma'} P^{II\sigma}_{ik}P^{JJ\sigma'}_{jl}-P^{IJ\sigma}_{il}P^{JI\sigma'}_{jk}](il|jk)
                          }
                         { \sum_{\sigma \sigma'} \sum_{ij} [\delta_{\sigma \sigma'}n^{II\sigma}_{ii} n^{JJ\sigma'}_{jj}-n^{IJ\sigma}_{ij}n^{JI\sigma'}_{ji}]
                         }.
\label{EqS:KIJ}
\end{equation}

The effects of cross exchange interactions on band gaps are summarized in Table S1.
As shown in Table A1, the calculated band gaps with and without effects of $K$ are negligile.
For solids considered here, all computed gaps with $K$ are little bit smaller than the gap without $K$.

\section{Effects of choice of pseudopotentials}

In this work, we used two kinds of pseudopotentials: 
PseudoDoJo norm-conserving~\cite{Setten2018CPC} 
and GBRV ultrasoft~\cite{Garrity2014CMS} pseudopotentials. 
Table~\ref{Table_A2} shows the effects of choice of pseudopotentials. 
Almost all $s$ and $p$ electron systems considered here
are not affected by the choice. 
However, particularly, in the case of Zn compound, the discrepancies are quite large.

\onecolumngrid
\clearpage
\begin{center}
\textbf{\large Supplemental Material for ``Efficient First-Principles Approach with a Pseudohybrid Density Functional for Extended Hubbard Interactions''}
\end{center}
\begin{center} \vspace{-5pt}
{Sang-Hoon Lee and Young-Woo Son}\\
\emph{ Korea Institute for Advanced Study, Seoul 02455, Korea}
\end{center}
\setcounter{equation}{0}
\setcounter{figure}{0}
\setcounter{table}{0}
\setcounter{page}{1}
\setcounter{section}{0}
\setcounter{subsection}{0}

\maketitle
\makeatletter
\renewcommand{\thesection}{\arabic{section}}
\renewcommand{\thesubsection}{\thesection.\arabic{subsection}}
\renewcommand{\thesubsubsection}{\thesubsection.\arabic{subsubsection}}
\renewcommand{\theequation}{S\arabic{equation}}
\renewcommand{\thefigure}{S\arabic{figure}}
\renewcommand{\thetable}{S\arabic{table}}

In Figs. S1-S8, we displayed the band structures of selected bulk solids given in the main manuscript. In order to show the effects of inter-site interactions V on the band structures, we present the results obtained by DFT-PBE (Blue lines), ACBN0 method (gray dashed lines) and our method (red solid lines). The energetic position of valence band maximum is set to zero for all figures.

\begin{figure} [h]
	\begin{center}
		\includegraphics[width=0.6\columnwidth]{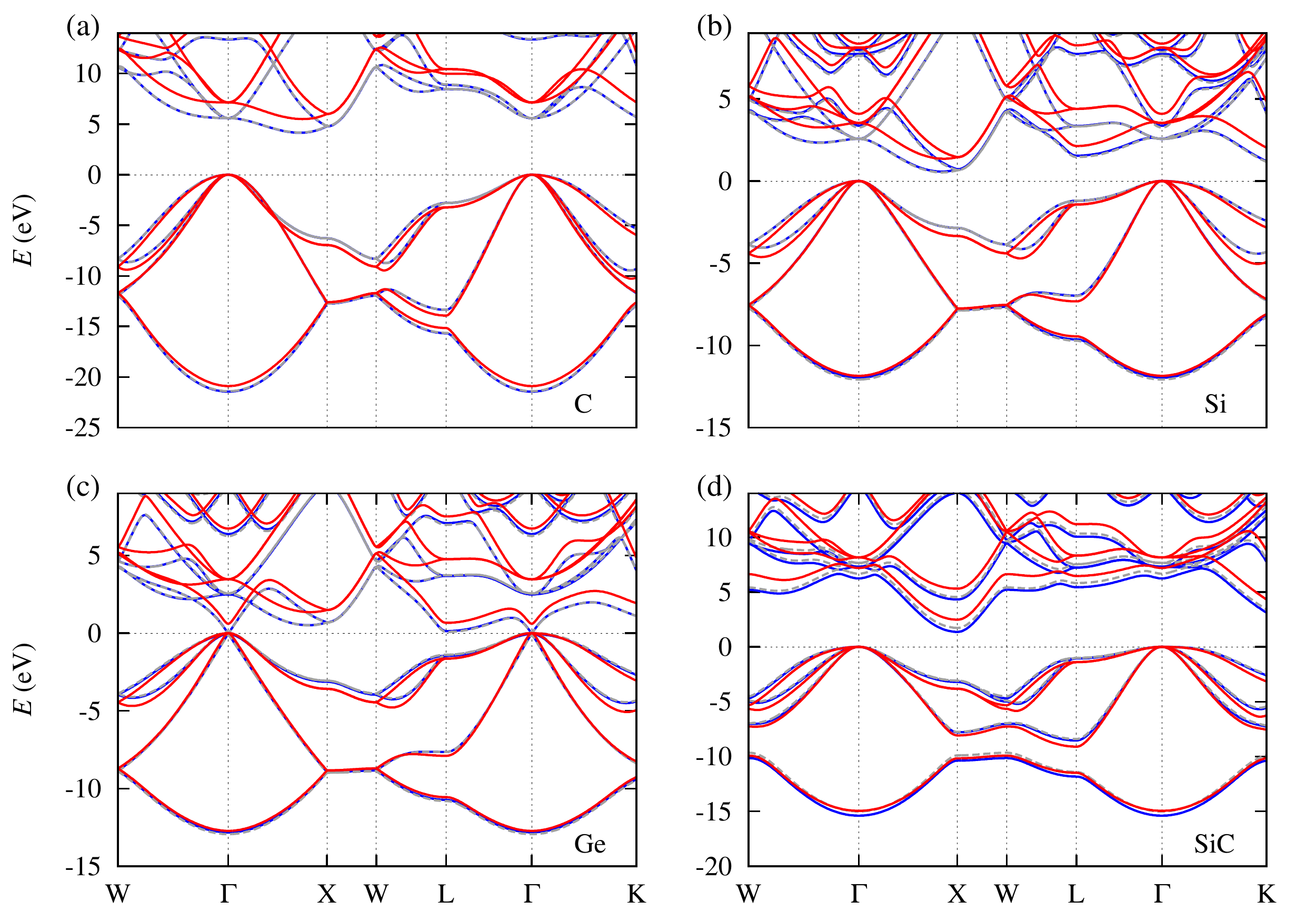}
	\end{center}
	\caption{Band structures of (a) C, (b) Si, (c) Ge, and (d) SiC}
\end{figure}

\begin{figure}[h]
	\begin{center}
		\includegraphics[width=0.6\columnwidth]{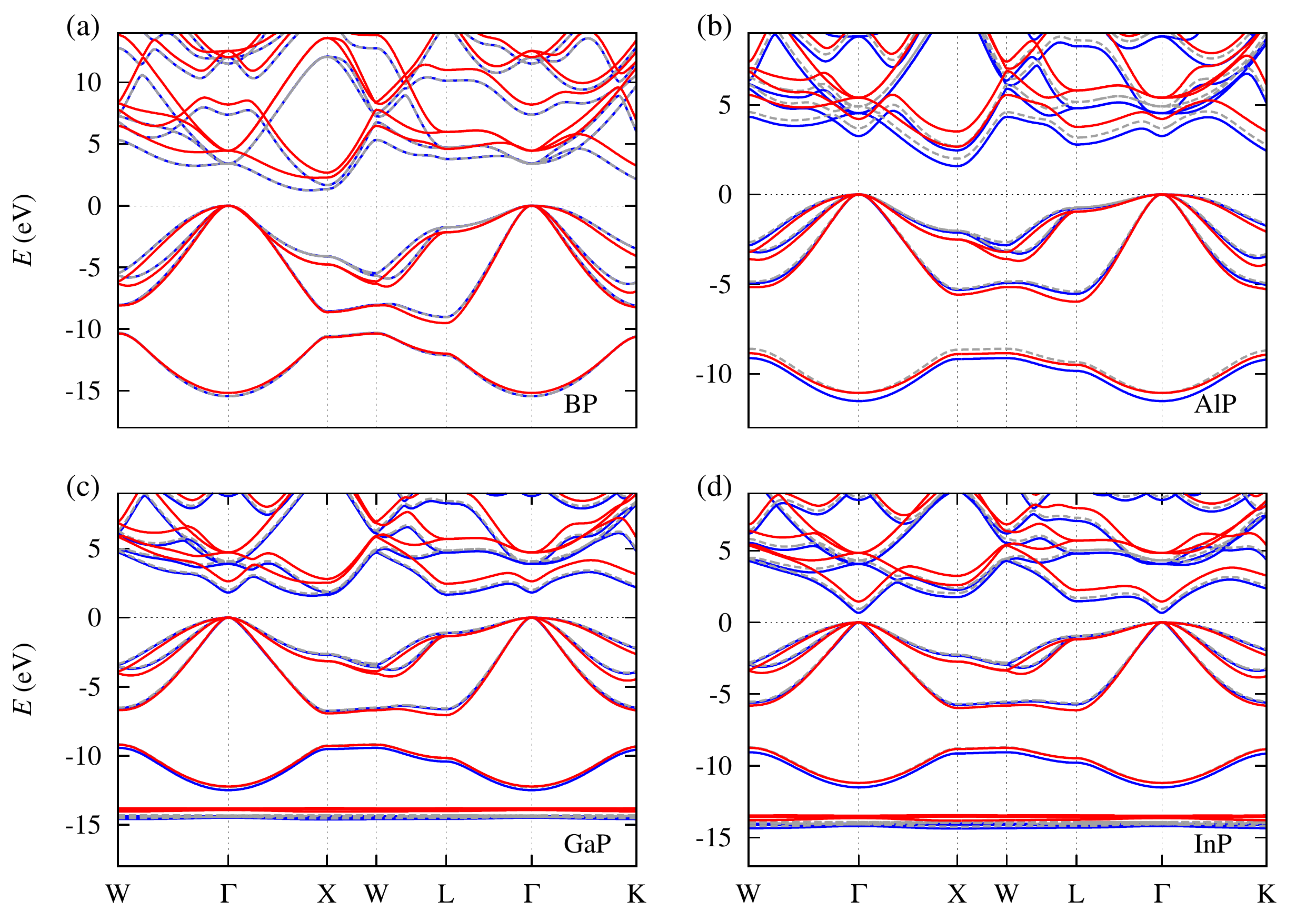}
	\end{center}
	\caption{Band structures of (a) BP, (b) AlP, (c) GaP, and (d) InP}
\end{figure}

\begin{figure}[h]
	\begin{center}
		\includegraphics[width=0.6\columnwidth]{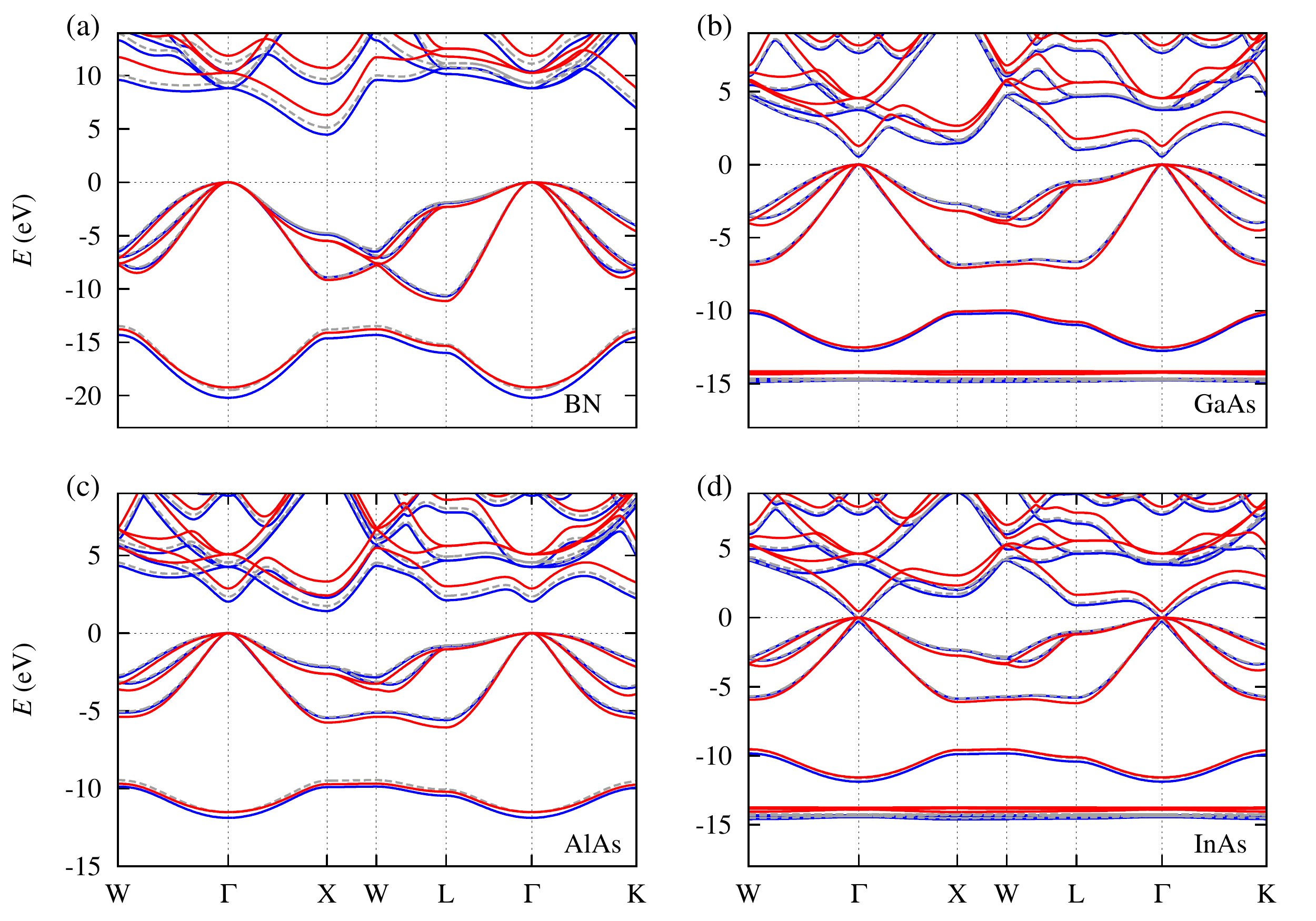}
	\end{center}
	\caption{Band structures of (a) BN, (b) GaAs, (c) AlAs, and (d) InAs}
\end{figure}

\begin{figure}[h]
	\begin{center}
		\includegraphics[width=0.6\columnwidth]{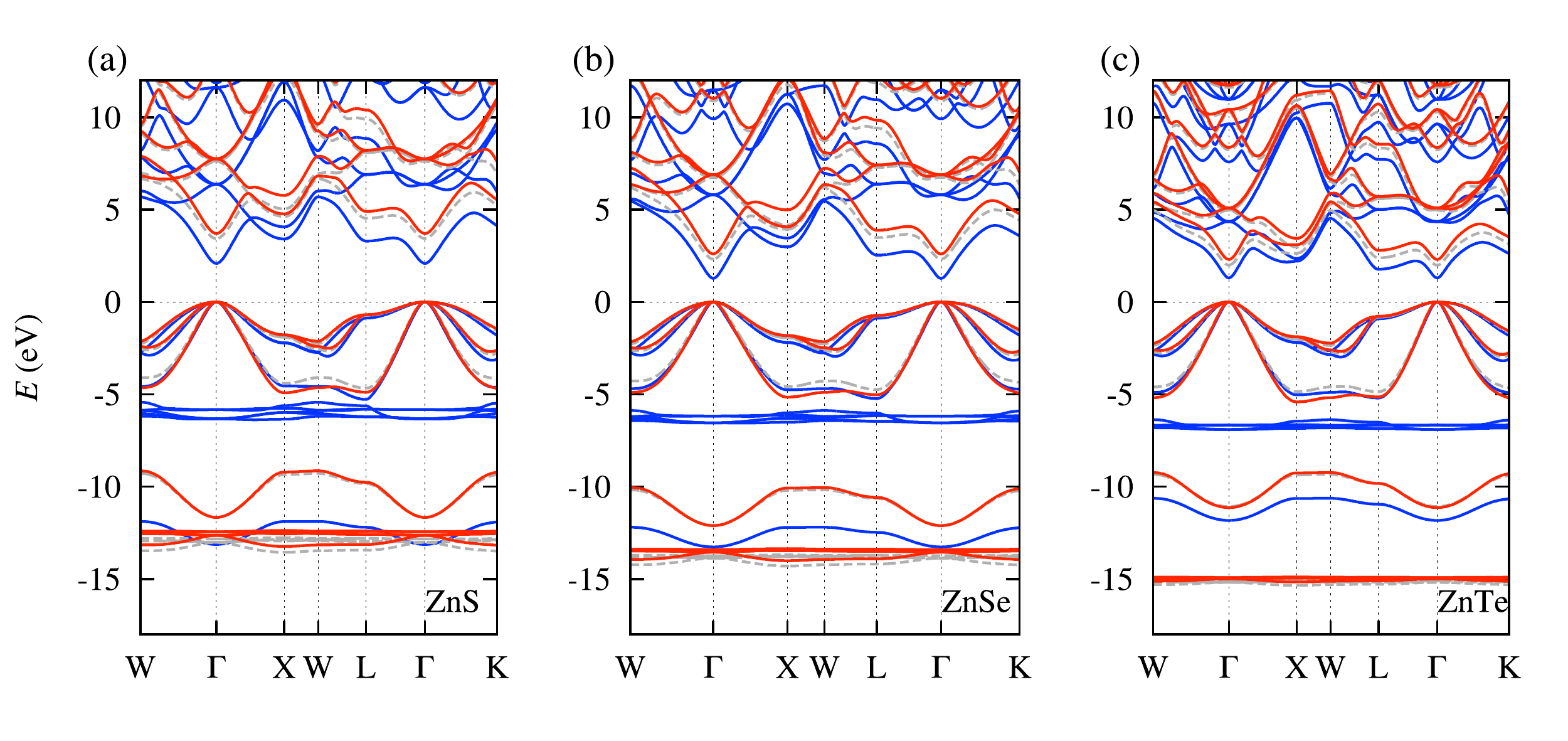}
	\end{center}
	\caption{Band structures of (a) ZnS, (b) ZnSe, and (c) ZnTe}
\end{figure}

\begin{figure}[h]
	\begin{center}
		\includegraphics[width=0.6\columnwidth]{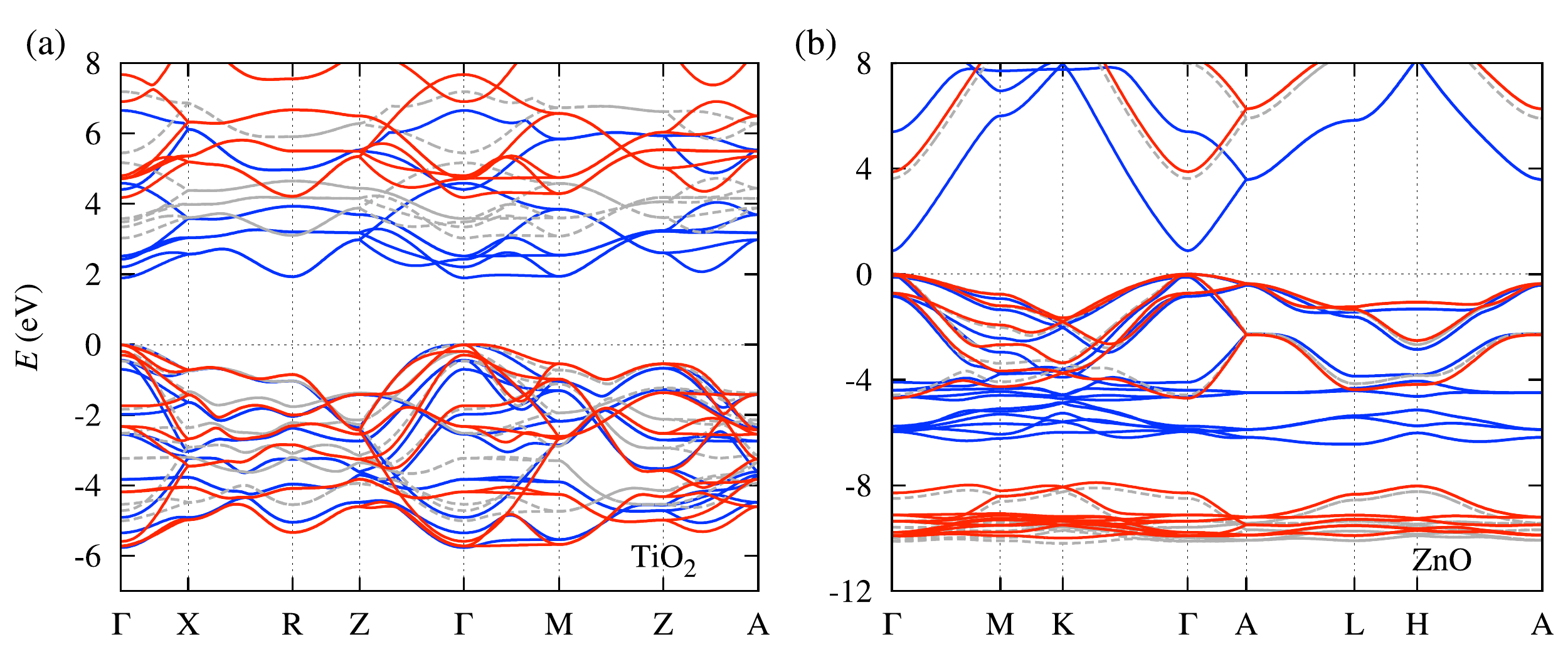}
	\end{center}
	\caption{Band structures of (a) Ti$\textrm{O}_2$, and (b) ZnO}
\end{figure}

\begin{figure}[h]
	\begin{center}
		\includegraphics[width=0.6\columnwidth]{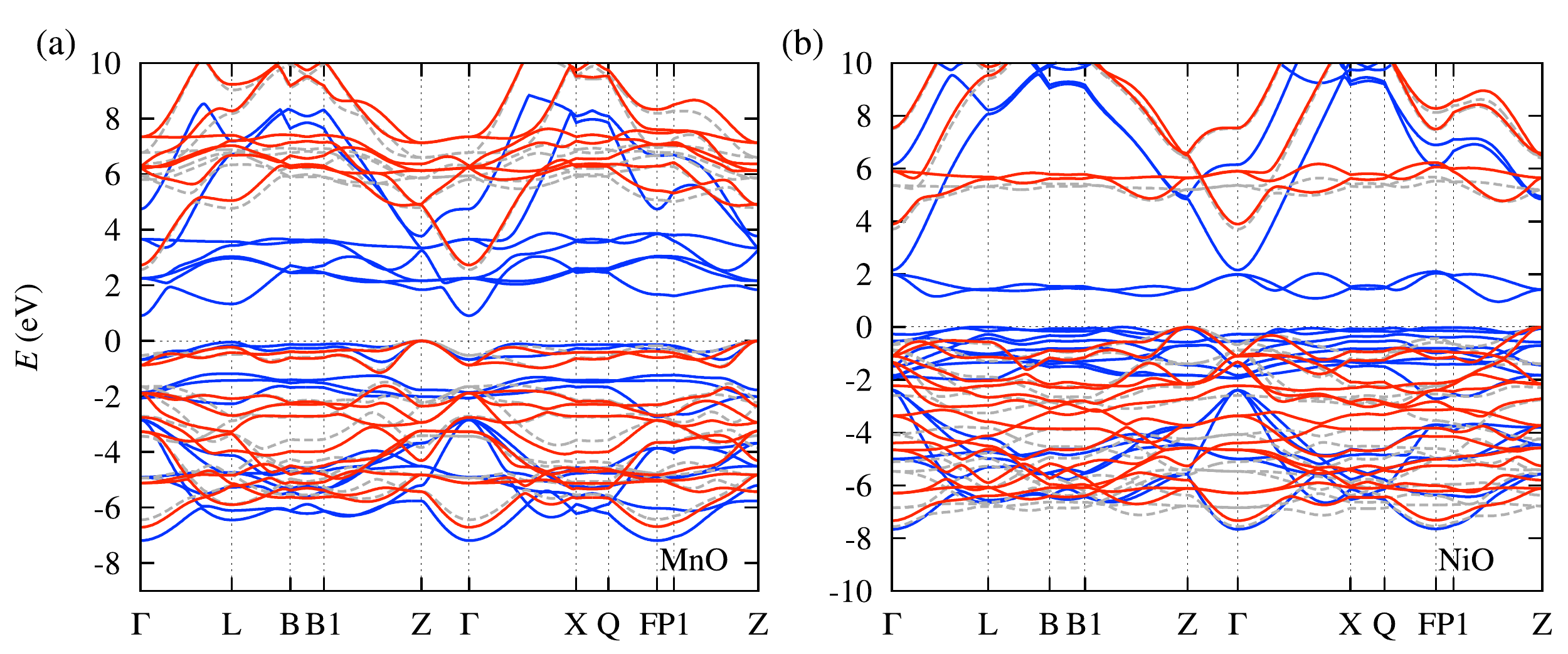}
	\end{center}
	\caption{Band structures of (a) MnO, and (b) NiO}
\end{figure}

\begin{figure}[h]
	\begin{center}
		\includegraphics[width=0.6\columnwidth]{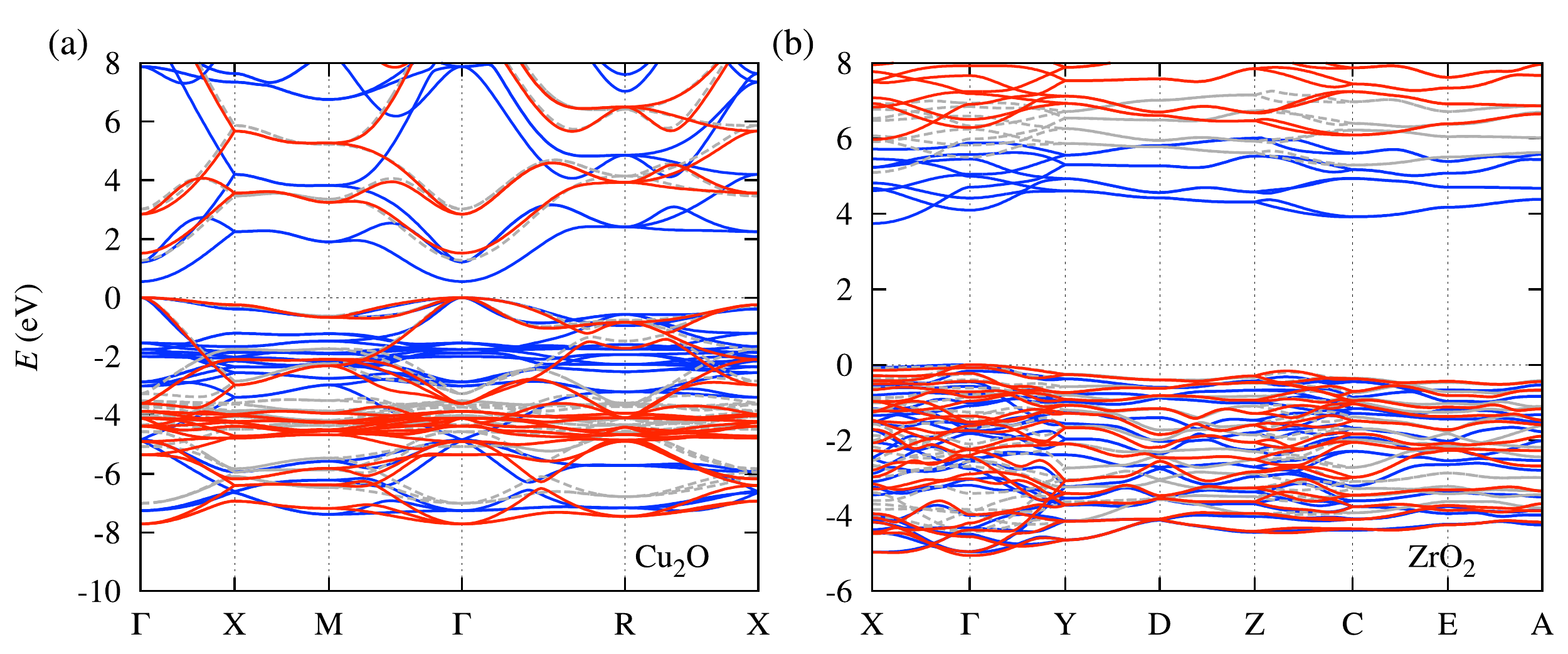}
	\end{center}
	\caption{Band structures of (a) $\textrm{Cu}_2$O, and (b) ZrO$_2$}
\end{figure}

\begin{figure}[h]
	\begin{center}
		\includegraphics[width=0.6\columnwidth]{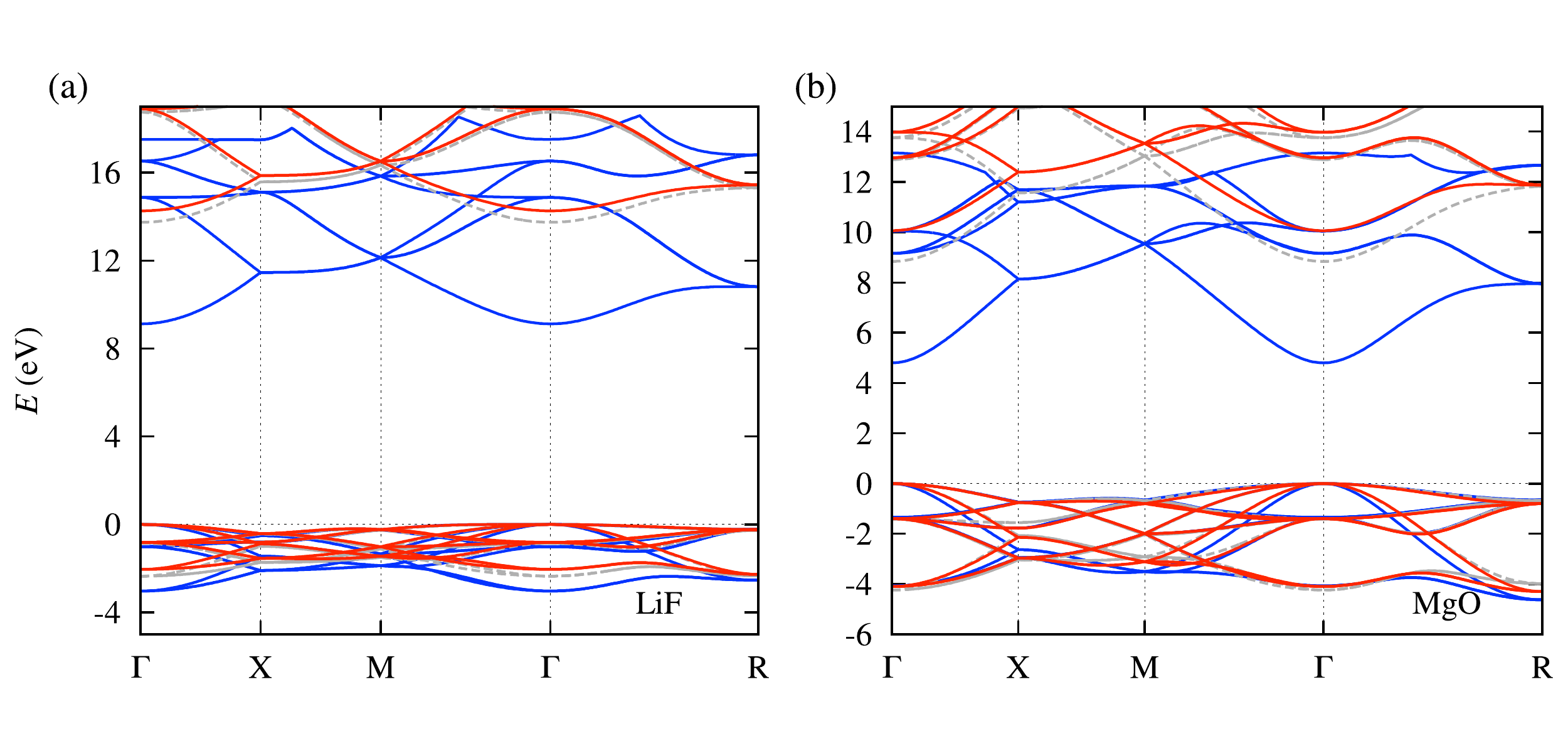}
	\end{center}
	\caption{Band structures of (a) LiF, and (b) MgO}
\end{figure}

\end{document}